\documentclass[a4paper,fleqn,usenatbib]{mnras}

\usepackage{mathptmx}

\usepackage[T1]{fontenc}
\usepackage{ae,aecompl,pdflscape}

\usepackage{graphicx}	
\usepackage{amsmath}	
\usepackage{amssymb}	
\usepackage{subfig}
\usepackage{float}
\usepackage{epstopdf}

\def\pb{Pa$\beta$}
\def\br{Br$\gamma$}

\def\feii{[Fe\,{\sc ii}]}

\def\oiii{[O\,{\sc iii}]}
\def\oiiil{[O\,{\sc iii}]$\lambda\,5007$}

\def\hml{H$_2$$\lambda 2.12\mu m$}
\def\p1{Paper~I}

\def\kms {$\rm km\,s^{-1}$}

\title[Gemini NIFS survey of feeding and feedback processes in AGN]{Gemini NIFS survey of feeding and feedback processes in nearby Active Galaxies: II -The sample and surface mass density profiles}
\author[Riffel et al.]{R. A. Riffel$^{1}$\thanks{E-mail: rogemar@ufsm.br}, T. Storchi-Bergamann$^2$, R. Riffel$^2$, R. Davies$^3$, M. Bianchin$^1$,
\newauthor  M. R. Diniz$^1$, A. J. Sch\"onell$^{2,4}$,  L. Burtscher$^5$, M. Crenshaw$^6$, T. C. Fischer$^7$, 
\newauthor L. G. Dahmer-Hahn$^2$, N. Z. Dametto$^2$, D. Rosario$^8$\\
$^{1}$ Departamento de F\'\i sica, CCNE, Universidade Federal de Santa Maria, 97105-900, Santa Maria, RS, Brazil \\ 
$^{2}$ Departamento de Astronomia, IF, Universidade Federal do Rio Grande do Sul, CP 15051, 91501-970, Porto Alegre, RS, Brazil \\
$^{3}$ Max-Planck-Institut f\"ur extraterrestrische Physik, Postfach 1312, D-85741, Garching, Germany\\
$^4$   Instituto Federal de Educa\c c\~ao, Ci\^encia e Tecnologia Farroupilha, BR287, km 360, Estrada do Chapad\~ao, 97760-000, Jaguari - RS, Brazil\\
$^{5}$ Leiden Observatory, Leiden University, PO Box 9513, 2300 RA Leiden, The Netherlands\\
$^6$   Department of Physics and Astronomy, Georgia State University, Astronomy Offices, 25 Park Place, Suite 605, Atlanta, GA 30303, USA \\
$^7$   Astrophysics Science Division, Goddard Space Flight Center, Code 665, Greenbelt, MD 20771, USA. \\
$^8$   Department of Physics, Durham University, South Road, Durham DH1 3LE, UK
}

\date{Accepted XXX. Received YYY; in original form ZZZ}

\pubyear{2017}

\begin{document}
\label{firstpage}
\pagerange{\pageref{firstpage}--\pageref{lastpage}}
\maketitle

\begin{abstract}

We present and characterize a sample of 20 nearby Seyfert galaxies selected for having BAT 14--195\,keV luminosities $L_X \ge 10^{41.5}$ ergs\,s$^{-1}$, redshift $z\le$0.015, being accessible for observations with the Gemini Near-Infrared Field Spectrograph (NIFS) and showing extended [OIII]$\lambda$5007 emission. Our goal is to study  Active Galactic Nuclei (AGN) feeding and feedback processes from near-infrared integral-field spectra, that include both ionized (H\,{\sc ii})  and hot molecular (H$_2$) emission. This sample is complemented by other 9 Seyfert galaxies previously observed with NIFS. We show that the host galaxy properties (absolute magnitudes $M_B$, $M_H$, central stellar velocity dispersion and axial ratio) show a similar distribution to those of the 69 BAT AGN. For the 20 galaxies already observed, we present surface mass density ($\Sigma$) profiles for H\,{\sc ii} and H$_2$ in their inner $\sim$500 \,pc, showing that H\,{\sc ii} emission presents a steeper radial gradient than H$_2$. This can be attributed to the different excitation mechanisms: ionization by AGN radiation for H\,{\sc ii} and   heating by X-rays for H$_2$. The mean surface mass densities are in the range ($0.2\le \Sigma_{HII} \le 35.9$)\,M$_\odot$\,pc$^{-2}$,  and ($0.2\le \Sigma_{H2} \le 13.9$)$\times10^{-3}$\,M$_\odot$\,pc$^{-2}$, while the ratios between the H\,{\sc ii} and H$_2$ masses range between $\sim$200  to 8000. The sample presented here will be used in future papers to map AGN gas excitation and kinematics, providing a census of the mass inflow and outflow rates and power as well as their relation with the AGN luminosity.

\end{abstract}

\begin{keywords}
galaxies: active -- galaxies: nuclei -- infrared: galaxies
\end{keywords}



\section{Introduction}


The co-evolution of Active Galactic Nuclei (AGN) and galaxies is now an accepted paradigm that permeates recent reviews \citep{kormendy13,heckman14}. But the conclusions put forth in these reviews are mostly based on surveys of integrated galaxy properties, and the feeding and feedback processes that lead to the co-evolution have been implemented in models in a simplistic way \cite{somerville08,springel05,croton06}. This is due to the  lack of observational constraints from spatially resolved studies. Physical motivated models \cite{hopkins10} show that the relevant feeding processes occur within the inner kiloparsec, that can only be resolved in nearby galaxies. The large quantities of dust in the inner kiloparsec of AGN, estimated to range from 10$^5$ to 10$^7$ M$_\odot$ \citep{simoeslopes07,martini13,audibert17} and the associated large content of molecular gas (10$^7$ to 10$^9$ M$_\odot$) points to the importance of looking for signatures of the feeding in the molecular gas within the nuclear region. Recently it has also been argued that the feedback in the form of massive outflows is also dominated by molecular gas \citep{sakamoto10,aalto12,veilleux13}, at least in LIRGS or ULIRGS (Ultra Luminous Infrared Galaxies). 

The co-evolution scenario, and the feeding of gas to the inner kiloparsec of galaxies when they are in the active phase, implies that the galaxy bulge grows in consonance with the SMBH. Since the early studies of Terlevich and collaborators \citep[e.g.][]{terlevich90}, it has been argued that the excess blue light and dilution of the absorption features of the nuclear spectra of active galaxies was due to young stars. Subsequent long-slit studies \citep{sb00,cid04,davies07,kauffmann09} have found an excess contribution of young to intermediate age stars to the stellar population in the inner kiloparsec of active galaxies when compared to non-active ones. This result has led to the proposition of an evolutionary scenario \citep{sb01,davies07, hopkins12}, in which the gas inflow to the nuclear region first triggers star formation in the circumnuclear region, and is then followed by the ignition of the nuclear activity.

Observational constraints for the feeding and feedback processes can be obtained via spatially resolved studies of nearby active galaxies using integral field spectroscopy (IFS). The radiation from the AGN heats and ionizes the surrounding gas in the galaxy up to hundreds of pc (and even kpc) scales. The heating excites rotational and vibrational states of the H$_2$ molecule that then emits in the near-IR, and the AGN radiation ionizes the gas that, in turn, emits permitted and forbidden lines that can be used to probe the ionized gas kinematics and excitation. Emission from both the molecular and ionized gas phases can be observed in the near-IR domain, where the effects of dust extinction is minimized. In the near-infrared, IFS at 10 meter class telescopes has been used to probe the feeding and feedback processes in nearby active galaxies, by  mapping and modeling the molecular and ionized gas kinematics in the inner kiloparsec of active galaxies -- on 10--100\,pc scales --  leading to insights on both the feeding and feedback mechanisms. For high signal-to-noise ratio in the continuum, the stellar  kinematics as well as the age distribution of the stellar population have also been mapped. So far, these studies show that (i)  Emission from molecular (H$_2$) and ionized gases present distinct flux distributions and kinematics.  The H$_2$ emission is distributed  all around the nucleus, seems to be located in the plane of the galaxy, shows low velocity dispersion ($<$100\,km\,s$^{-1}$) and is dominated by rotational motion. In few cases, a very steep rotation curve is observed, suggesting the presence of compact molecular disks \cite{mrk1066-kin,mrk766,hicks13,mazzalay14}.  In a number of cases, streaming motions towards the central regions were mapped along nuclear spiral arms with estimated inflow rates in total molecular gas ranging from a few tenths to a few solar masses per year \citep{n4051,mrk79,davies09,ms09,diniz15}. 
(ii) The ionized gas emission is more collimated and shows higher velocity dispersion ($>$\,100\,km\,s$^{-1}$) than the molecular gas, seems to extend to high latitudes and its kinematics comprises both rotation and outflow \citep[e.g.][]{eso428,mrk1066-exc,ms11,mrk79,barbosa14,sb10}. 
(iii) Only for a few cases, the study of stellar population was done using near-IR IFS. These works show the presence of young to intermediate age ($\sim$10$^8$ yr) stars, usually in $\sim$\,100\,pc rings \citep[e.g.][]{mrk1066-pop,mrk1157-pop,sb12}, that correlate with rings of low velocity dispersion. This correlation has been interpreted as being  
a signature of the co-evolution of the bulge and SMBH: as the estimated mass inflow rates are $\sim$\,3 orders of magnitude larger than the accretion rate to the AGN, most of the molecular gas that is accumulated in the nuclear regions of AGNs is forming new stars in the inner few hundred parsecs of the galaxy, leading to the growth of the bulge.

Most of the results summarized above were obtained by studying individual galaxies, selected using distinct criteria, and a study of a well-defined, comprehensive sample is of fundamental importance to understand the relation among AGN feeding, feedback and galaxy evolution \citep[e.g.][]{davies17}. In the present work, we describe a sample of nearby active galaxies that are being observed with the Gemini Near-Infrared Integral Field Spectrograph (NIFS). Our aim with these observations is to study the details of the inner few hundreds of parsecs of AGNs and better constrain the feeding and feedback processes. This is the second paper of a series in which we will be mapping the gas excitation and kinematics, as well as the stellar population characteristics and kinematics. In the first paper \citep{stel_llp}, we have presented and discussed stellar kinematics measurements for 16 galaxies of the sample  and in forthcoming papers we will analyse the emission-line flux distributions, gas kinematics and map the stellar populations. This paper is organized as follows: Section~\ref{defsample} describes the selection criteria of the sample, the instrument configuration, observations, data reduction and compare nuclear and large scale properties of the galaxies. In section~\ref{massdens} we present and discuss measurements of the molecular and ionized gas masses and surface densities for the galaxies already observed and Section~\ref{feeding} discusses the implications of the derived amount of gas to the AGN feeding process and star formation.  Finally, section~\ref{conclusions} presents the conclusions of this work.

\section{Definition of a sample and observations} \label{defsample}

\subsection{The sample} \label{sample}
In order to select out an AGN sample, we used the  Swift-BAT 60-month catalogue \citep{ajello12}, and  selected nearby galaxies with 14--195\,keV luminosities $L_X \ge 10^{41.5}$ ergs\,s$^{-1}$ and redshift $z\le$0.015. The hard (14--195\,keV) band of the Swift-BAT survey measures direct emission from the AGN rather than scattered or re-processed emission, and is much less sensitive to obscuration in the line-of-sight than soft X-ray or optical wavelengths, allowing a selection based only on the AGN properties. \citet{davies15} describe a southern hemisphere sample selected in a similar way and discuss its rationale for studying AGN feeding and feedback processes \citep[see also, ][]{davies17}. Although their sample includes brighter and closer galaxies then ours, being composed by galaxies with log\,$L_{\rm X}$$=42.4 - 43.7$ and $z<0.01$.

As additional criteria, the object must be accessible for Gemini NIFS ($-30^\circ< \delta < 73^\circ$) and its nucleus being bright/pointy enough to guide the observations or with natural guide stars available in the field. Finally, we only have included in the sample galaxies already previously observed in the optical and with extended \oiii$\lambda5007$ emission  available in the literature. We have used this constraint in order to ensure that we will have extended gas emission to allow spatially resolve its kinematics  and look for possible inflows and outflows. From our previous experience, a galaxy that shows extended \oiii\  emission will also have a similarly extended \feii\  or \pb\ emission. Table~\ref{sampletab} presents the resulting sample, which is composed of 20 galaxies. In addition, we included 9 galaxies observed with NIFS by our group in previous works (shown below the horizontal line in Table~\ref{sampletab}). These aditional galaxies may be used as a complementary sample in forthcoming works.

\begin{figure}
\includegraphics[scale=0.45]{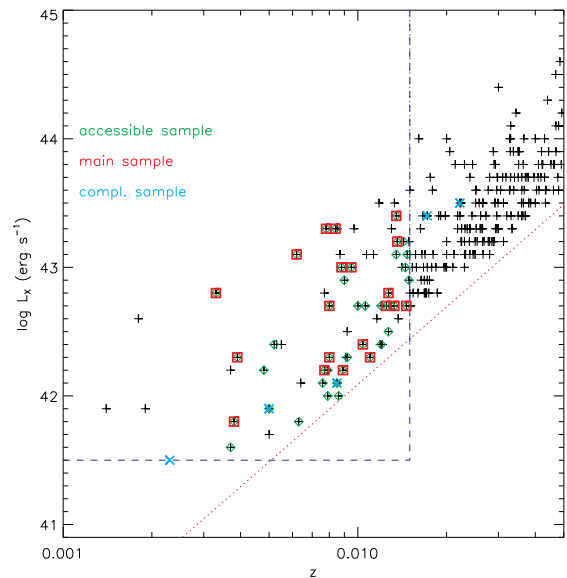}
\caption{Plot of $L_X$ vs. $z$ for the galaxies of our sample. Black crosses show all objects (257) with log\,$L_X > 41.5$ at the  Swift BAT 60-month catalogue, green diamonds represent objects (43) accessible by NIFS ($-30^\circ< \delta < 73^\circ$), red squares represent our main sample  (20) and cyan crosses are objects from our complementary sample detected in Swift BAT. All points at $z< 0.015$ make up what we call ``the restricted BAT sample", composed by 69 galaxies. The red dotted line shows the detection limit of Swift and the dashed lines show the limits in  $L_X$ and $z$  used the NIFS sample ($L_X \ge 10^{41.5}$ ergs\,s$^{-1}$ and $z\le$0.015). }
\label{lxz}
\end{figure}

Figure~\ref{lxz} shows a plot of $L_X$ vs. $z$ for  all Swift BAT AGN with $z\leq0.05$ and accessible to Gemini North ($-30^0 < \delta < 73^0$).
 Green diamonds show the galaxies accessible to Gemini North that satisfy the following criteria: $L_X \ge 10^{41.5}$ ergs\,s$^{-1}$ and $z\le$0.015, while the red squares show our main sample (objects that satisfy all the requirements above). The cyan $\times$ symbols show the objects of the complementary sample detected in the Swift-BAT 60-month catalogue. The red dotted line shows the detection limit of the Swift 60-month catalogue and the vertical and horizontal lines show the $L_X$ and $z$ cuts used to define our sample, respectively.


\begin{table*} 
\centering
\caption{The sample. (1) Galaxy name; (2) Redshift; (3) Morphological classification; (4) Nuclear Activity (from quoted in NED), (5) Swift 14-195 keV luminosity, (6) [O\,{\sc iii}]$\lambda5007$ luminosity in units of ergs\,s$^{-1}$, (7) reference for the \oiii\ luminosity. Table~\ref{sampleobs} list the galaxies already observed.}
\vspace{0.3cm}
\begin{tabular}{l c c c c c c}
\hline
       (1)         &   (2)     &    (3)            &  (4)   &    (5)    & (6)	     & (7)\\
Galaxy 	           &$z$      & Hubble Type  &Nuc. Act.   &log($L_{\rm X}$) &log($L_{\rm OIII}$) & Ref.	 \\
\hline								    		   		       
\multicolumn{7}{c}{Main sample}\\
\hline 
         NGC788   &  0.014    &SA0/a?(s)	  &   Sy2  &  43.20   &    41.06       &  a    \\
        NGC1068   &  0.004    &(R)SA(rs)b	  &   Sy2  &  41.80   &    41.53       &  b	 \\
        NGC1125   &  0.011    &(R')SB0/a?(r)	  &   Sy2  &  42.30   &    39.69       &  c		    \\
        NGC1194   &  0.013    &SA0$^+$? 	  &   Sy1  &  42.70   &    39.60       &  b		 \\
        NGC2110   &  0.008    &SAB0$^-$ 	  &   Sy2  &  43.30   &    40.64       &  a   \\
           Mrk3   &  0.014    &S0?		  &   Sy2  &  43.40   &    41.83       &  b	\\
        NGC2992   &  0.008    &Sa pec		  &   Sy2  &  42.20   &    41.42       &  a	    \\
        NGC3035   &  0.015    &SB(rs)bc 	  &   Sy1  &  42.70   &    39.83       &  c		   \\
        NGC3081   &  0.008    &(R)SAB0/a(r)	  &   Sy2  &  42.70   &    41.58       &  a	      \\
        NGC3227   &  0.004    &SAB(s)a pec	  & Sy1.5  &  42.30   &    40.84       &  a	     \\
        NGC3393   &  0.013    &(R')SB(rs)a?	  &   Sy2  &  42.70   &    41.58       &  b		\\
        NGC3516   &  0.009    &(R)SB0$^0$?(s)	  & Sy1.5  &  43.00   &    41.02       &  b		 \\
        NGC3786   &  0.009    &SAB(rs)a pec	  & Sy1.8  &  42.20   &    40.59       &  a		\\
        NGC4151   &  0.003    &(R')SAB(rs)ab?	  & Sy1.5  &  42.80   &    42.19       &  a	 \\
        NGC4235   &  0.008    &SA(s)a edge-on	  &   Sy1  &  42.30   &    39.31       &  a		\\
         Mrk766   &  0.013    &(R')SB(s)a?	  & Sy1.5  &  42.80   &    41.10       &  b	  \\
        NGC4388   &  0.008    &SA(s)b? edge-on    &   Sy2  &  43.30   &    41.26       &  b		 \\
        NGC4939   &  0.010    &SA(s)bc  	  &   Sy1  &  42.40   &    40.64       &  c		     \\
        NGC5506   &  0.006    &Sa pec edge-on	  & Sy1.9  &  43.10   &    40.97       &  a		     \\
        NGC5728   &  0.009    &SAB(r)a? 	  &   Sy2  &  43.00   &    41.47       &  a		     \\
\hline
\multicolumn{7}{c}{Complementary Sample} \\
\hline  															        
NGC1052   &  0.005    &E4		   &  Sy2  &  41.90    &      --       &    --  \\
NGC4051   & 0.002     &SAB(rs)bc	   &   Sy1 & 41.50     &   --	       & --  \\
NGC5548   &0.017      &(R')SA0/a(s)	  &   Sy1 & 43.40      &  41.37        & b \\
NGC5899   &  0.009    &SAB(rs)c 	   &  Sy2  &  42.10    &      --       &    --        \\
NGC5929   & 0.008     &Sab? pec 	   &  Sy2  & --        &      --       &    --         \\
Mrk79     & 0.022     &SBb		   &   Sy1 &  43.50    & 41.58         &  b  \\
Mrk607    & 0.009     &Sa? edge-on	   &   Sy2 &  --       &      --       &    --         \\
Mrk1066   & 0.012     &(R)SB0$^+$(s)	   &  Sy2  &  --       &    --         & -- \\
Mrk1157   & 0.015     &(R')SB0/a	   &  Sy2  &  --      & --	       &--  \\
\hline									
\multicolumn{7}{l}{ References: a: \citet{wittle92}, b: \citet{schmitt03}, c: \citet{gu06};  }\\
\multicolumn{7}{l}{ d: \citet{noguchi10}; e:  \citet{zhu11}. }\\
\end{tabular}
  \label{sampletab}
\end{table*}

\begin{figure*}
\includegraphics[width=0.95\textwidth]{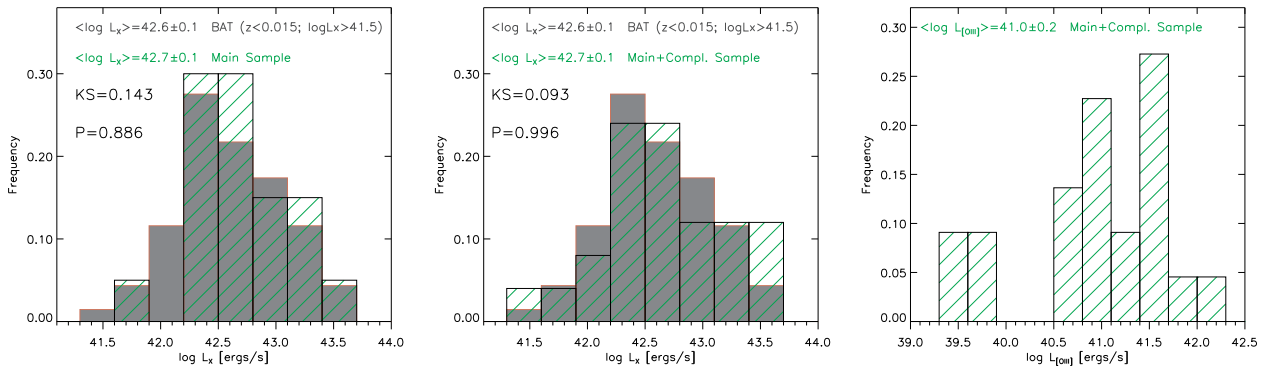}
\caption{Histograms for the distribution of X-ray and \oiiil\ luminosities of the galaxies of our sample. The left panel shows the distribution of log\,$L_{\rm X}$ of all galaxies with  $L_X \ge 10^{41.5}$ ergs\,s$^{-1}$ and $z\le$0.015 from the 60 month BAT catalogue (the ``restricted BAT" sample) in gray, with the distribution of our main sample overploted and crosshatched green histogram.  In the central panel, the complementary sample is included and the right panel shows the distribution of the \oiiil\ luminosities for our sample, including the two objects from the complementary sample with \oiii\ luminosities available. All histograms were constructed using a bin of log\,$L_{\rm X}=0.3$\,erg\,s$^{-1}$ and the mean values for each distribution are shown at the top of each panel. The results for the K-S statistical test ($KS$ and $P$) are shown  are shown for the first two panels.}
\label{hlx}
\end{figure*}

\begin{figure*}
\includegraphics[width=0.85\textwidth]{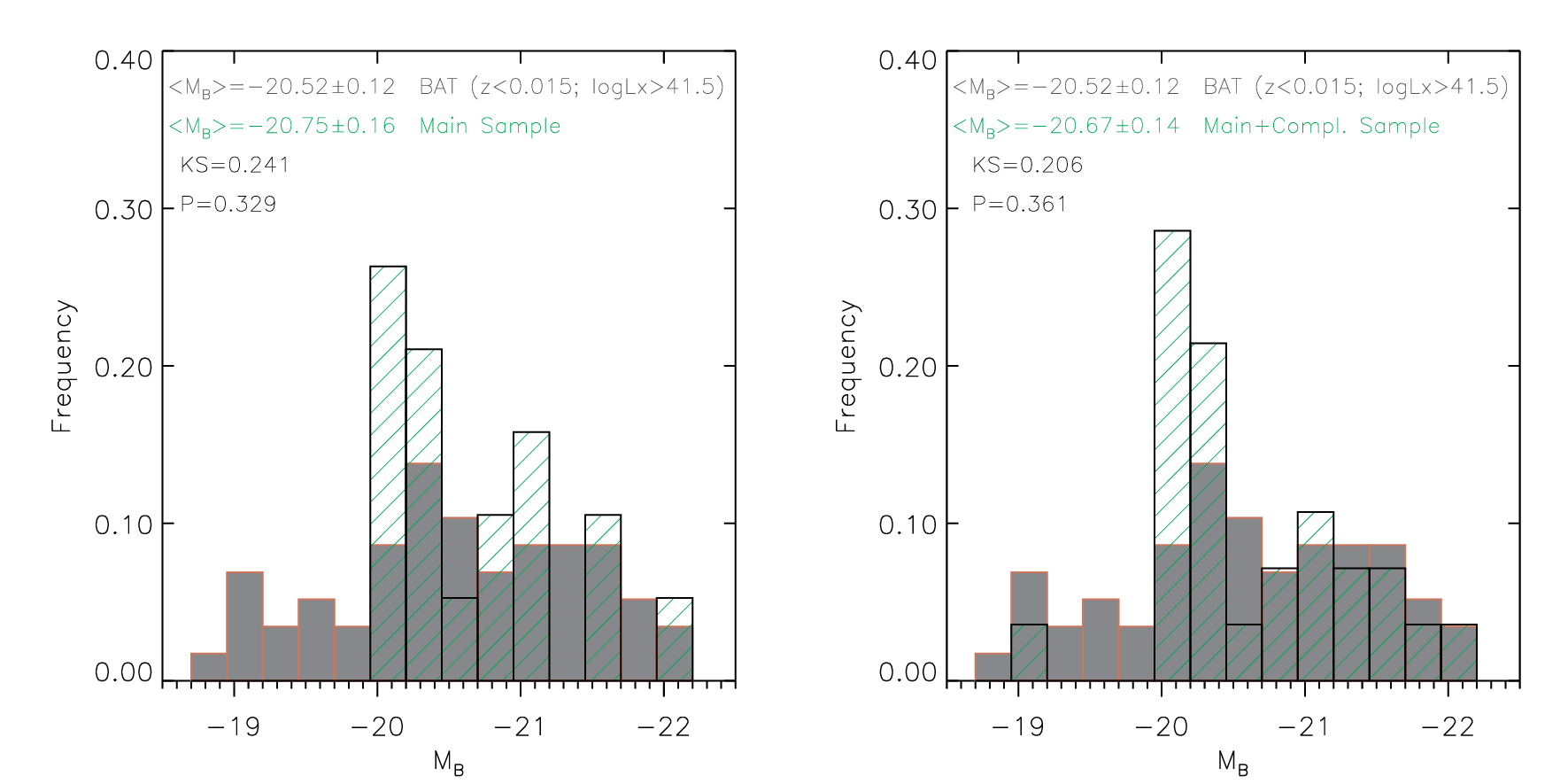}
\includegraphics[width=0.85\textwidth]{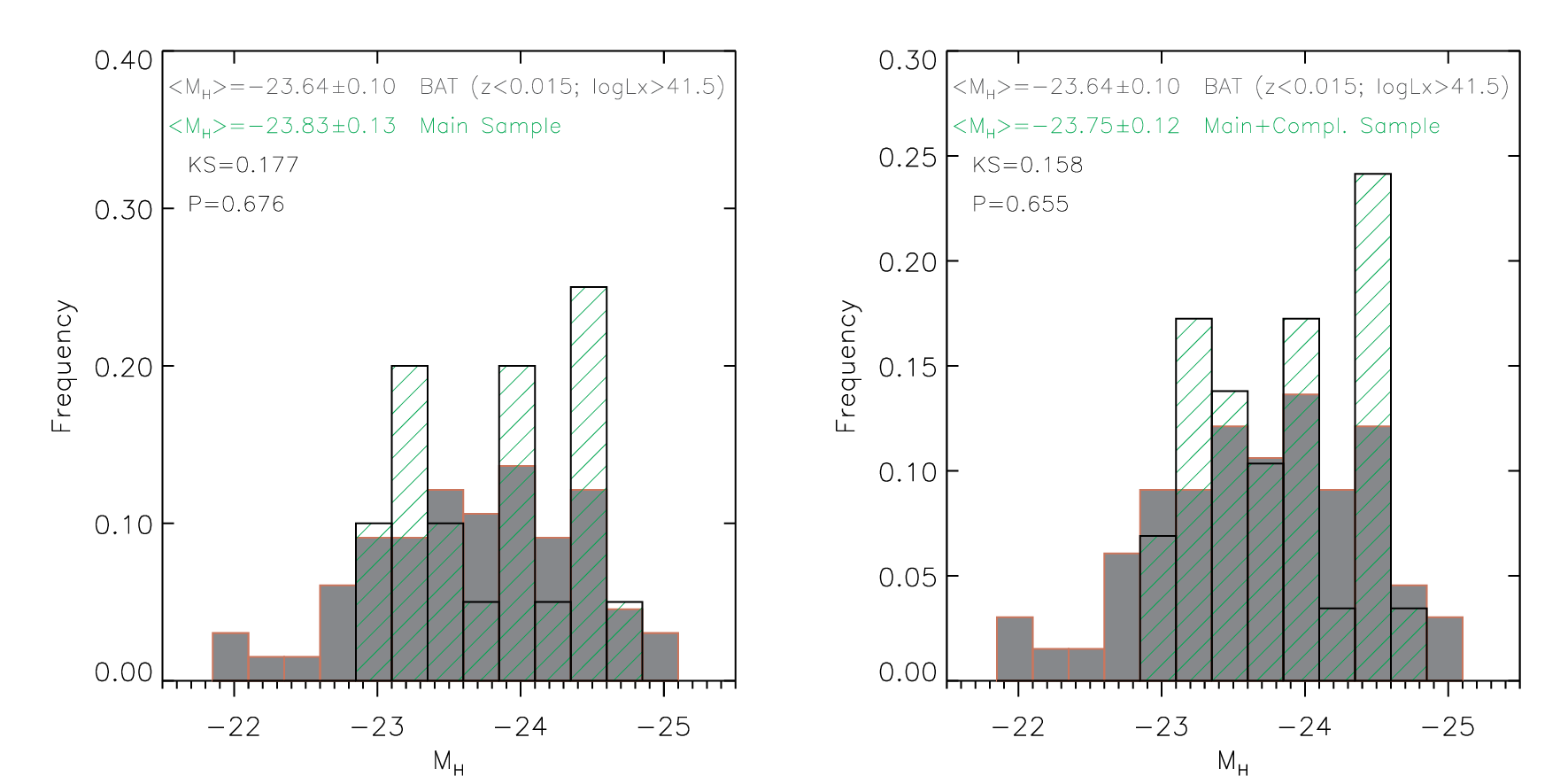}
\caption{Distribution of B (top) and H (bottom) band absolute magnitudes for the galaxies of the main sample (left) and main$+$complementary sample (right) in bins of 0.25 mag. The distribution of the  BAT sample is shown as the gray histogram. The results for the K-S statistical test ($KS$ and $P$) are shown  in each panel.}
\label{hmag}
\end{figure*}

\begin{figure*}
\includegraphics[width=0.85\textwidth]{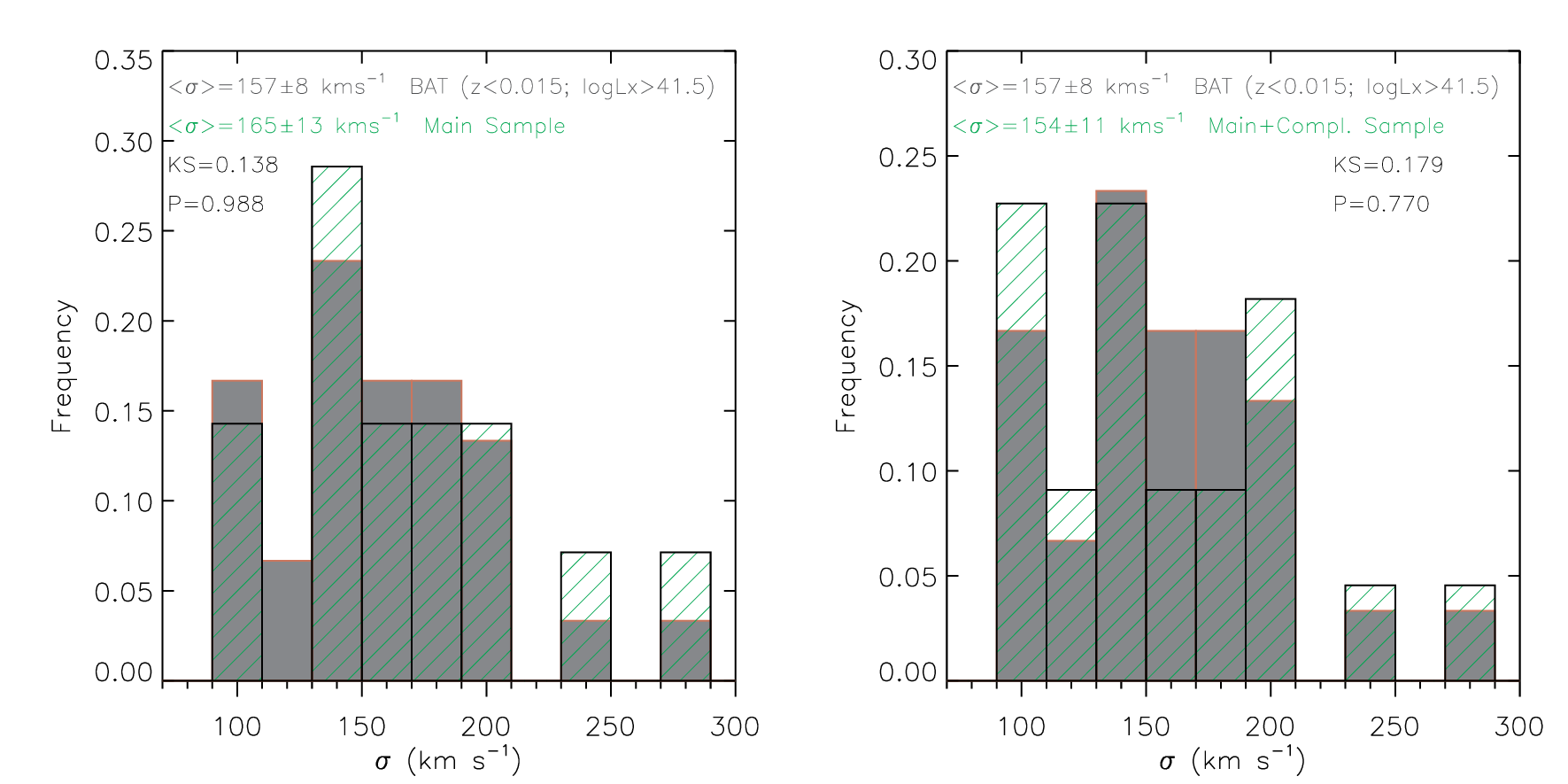}
\includegraphics[width=0.85\textwidth]{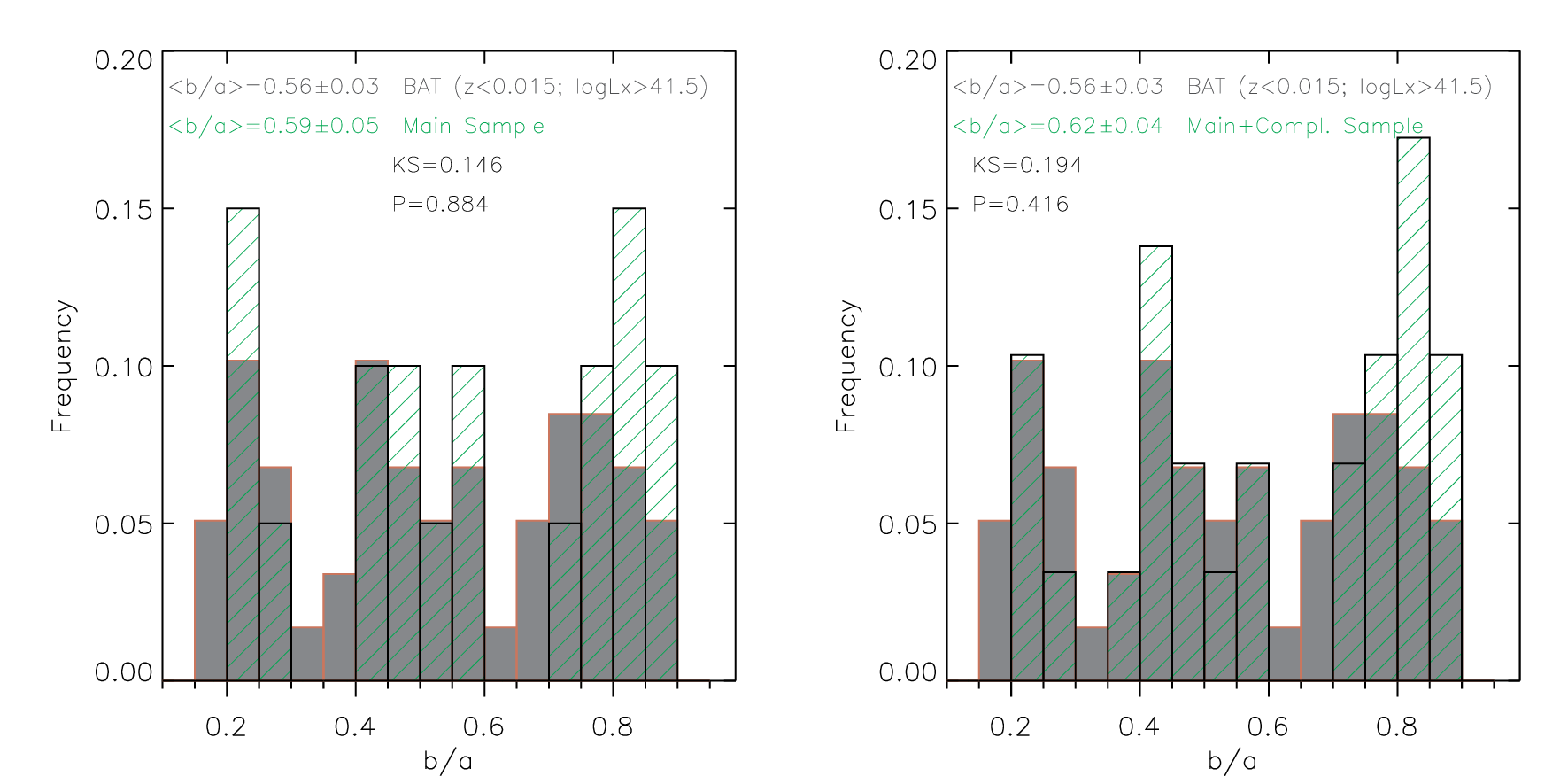}
\caption{Distribution of central stellar velocity dispersion (top) and axial ratio (bottom) for the galaxies of the main sample (left) and main$+$complementary sample (right). The distribution of the restricted BAT sample is shown in gray scale on the background. Bins of 20~km\,s$^{-1}$ are used in the histograms for the velocity dispersion and of 0.05 for the axial ratio. The results for the K-S statistical test ($KS$ and $P$) are shown  in each panel.}
\label{hkin}
\end{figure*}

\subsection{Characterization of the sample}

It is well known that hard X-ray emission is a good tracer of nuclear activity in galaxies, and thus a X-ray selected sample is representative of the population of AGN within the limited volume. However, besides the limits in X-ray luminosity and redshift, we included a constraint based on the detection of [O\,{\sc iii}]$\lambda5007$ emission line in order to increase the rate of detection of extended emission in near-IR lines, necessary to map the gas kinematics and flux distributions. In order to test if this additional criteria produces any bias on our sample, 
as compared to objects selected only on the basis of their X-ray emission, we compare the distribution of physical properties of the nucleus and host galaxies of the BAT sample (composed of galaxies with $L_X \ge 10^{41.5}$ ergs\,s$^{-1}$ and redshift $z\le$0.015) with the distributions of our main and complementary samples.  

The total number of galaxies in the 60 month BAT catalogue that follows the constraints above is 69 galaxies (hereafter we will call this sample as the ``restricted BAT sample"), while our main sample is composed of 20 objects, as shown in Table~\ref{sampletab}. In the left panel of Figure~\ref{hlx} we present a histogram for the distribution of $L_X$ of our main sample in bins of  log\,$L_{\rm X}=0.3$ crosshatched histogram, overlaid on the histogram for the restricted BAT sample, which is shown in gray. As can be observed in this plot, both samples show a very similar distribution with mean luminosities of $<{\rm log} L_{\rm X}> =42.6\,\pm\,0.1$ erg\,s$^{-1}$ and $<{\rm log} L_{\rm X}> =42.7\,\pm\,0.1$ erg\,s$^{-1}$ for the BAT and our main sample, respectively. We performed a Kolmogorov-Smirnov (K-S) statistic test to estimate the K-S confidence index ($KS$) and the probability of the two distributions being drawn from the same distribution ($P$). The resulting parameters are $KS=0.143$ and $P=0.886$, indicating that both restricted BAT and main samples have a probability of $\sim$89\,\% of being originated from the same distribution. Thus, the inclusion of the additional selection criteria of having extended [O\,{\sc iii}] emission already published and being observable with NIFS does not change significantly the distribution of the sample in terms of X-ray luminosities and our  main sample can be considered a representative sample of nearby AGNs within adopted constrains in X-ray luminosity and redshift. It is already well known that a close correlation between  the [O\,{\sc iii}] and hard X-ray luminosities is observed for AGNs \citep[e.g.][]{lamperti17} and that a better correlation is found if the sample is selected based on the X-ray luminosity than if it is drawn from [O\,{\sc iii}] luminosity \citep{heckman14}. As our sample is based on the X-ray luminosity, the similarity in the X-ray distribution is both samples is expected.

Five galaxies of the complementary sample have X-ray luminosities available in the 60 month BAT catalogue. Including these sources, the distribution of galaxies shows an extension to lower X-ray  luminosities as seen in the central panel of Fig.~\ref{hlx}, filling the low-luminosity ``gap" seen in the main sample, as only one galaxy of the complementary sample shows ${\rm log} L_{\rm X} >42.3$. However, the averaged luminosity does not change, as the complementary sample includes also two high luminosity objects (NGC\,5548 and Mrk\,79). The K-S test indicates that the inclusion of these sources makes the sample even more similar to the restricted BAT sample, with almost 100~\,\% of probability of both samples follow the same distribution in $L_{\rm X}$. Besides the 20 galaxies of our main sample, \oiiil\ luminosities are available for two galaxies of the complementary sample.  Our combined (main $+$ complementary) sample shows \oiiil\ luminosities in the range  $L_{[OIII]} = (0.2-155)\times10^{40}$\,erg\,s$^{-1}$, with a mean value of $<{\rm log} L_{\rm [OIII]}> =41.0\,\pm\,0.2$ erg\,s$^{-1}$.

We compiled physical properties of the host galaxies from the Hyperleda database \footnote{The Hyperleda database is available at $http://leda.univ-lyon1.fr/$} \citep{makarov14} and NED\footnote{NASA/IPAC  Extragalactic  Database available at $http://ned.ipac.caltech.edu/$ }. In figures~\ref{hmag} and \ref{hkin} we present histograms for the absolute B (top panels of Fig.~\ref{hmag}) and H magnitudes (bottom panels of Fig.~\ref{hmag}), the nuclear stellar velocity dispersion (top panels of Fig.~\ref{hkin}) and axial ratio (bottom panels of Fig.~\ref{hkin}). Both magnitudes correspond to apertures that include the the total emission of the host galaxy. The left panels of these figures show the distribution of these properties for the main sample, while the right panels show the same properties for the combined  sample. As in Fig.~\ref{hlx} the restricted BAT sample is shown as the gray histogram.

The B absolute magnitude $M_B$ was obtained from the Hyperleda database, and is available for 58 objects from the restricted BAT sample and for 28 galaxies of our sample, the only exception being NGC\,3035. The mean value of $M_B$ for our main sample ($M_B=-20.75\pm0.16$ mag) is similar to that of the BAT sample ($M_B=-20.52\pm0.12$ mag), but the distributions are somewhat distinct as the BAT sample includes more low luminosity galaxies with $M_B> -20$ mag.
The K-S test results gives a probability of $\sim$33\,\% that the main and restricted BAT samples follow the same distribution in $M_B$, while including the complementary sample, this probability increases to $\sim$36\,\%, being still small.

The total H absolute magnitude was obtained from the apparent H magnitudes from the The Two Micron All Sky Survey catalogue\footnote{Available at $http://vizier.u-strasbg.fr/viz-bin/VizieR$} \citep[2MASS,][]{2mass}. The H band is dominated by emission from the galaxy bulges and its luminosity can be used as a proxy for stellar mass of the galaxy \citep{davies15,davies17}. As for $M_B$, the distribution of the galaxies of our main sample is similar to that of the composite sample and the mean value of $M_H$ for both samples are very similar to that observed for the BAT sample.  However, for $M_H$ the K-S test indicates that there about 68\,\% of probability of both samples follow the same distribution. A similar $P$ value is found if we include the complementary sample.

In Figure~\ref{hkin} we show histograms for the distribution of the nuclear stellar velocity dispersion ($\sigma$ -- top panels) and axial ratio ($b/a$ -- bottom panels). The $\sigma$ values were obtained from the Hyperleda database and are standardized to an aperture of 0.595 $h^{-1}$\,kpc.
Measurements of $\sigma$ are available at Hyperleda database for 30 galaxies of the restricted BAT sample, 14 galaxies of the main sample and 8 objects of the complementary sample. The histograms for $\sigma$ were constructed using bins of 20\,km\,s$^{-1}$. As seen in Fig.~\ref{hkin} the distribution of $\sigma$ values for the main and restricted BAT samples are similar, with mean $\sigma$ values of $<\sigma>=165\pm13$\,km\,s$^{-1}$ and  $<\sigma>=157\pm8$\,km\,s$^{-1}$, respectively. By including the complementary sample, the fraction of objects with $\sigma\leq120$\,km\,s$^{-1}$ increases, while the mean $\sigma$ values are still consistent with that of the restricted BAT sample, as observed at the top-right panel of Fig.~\ref{hkin}. The K-S test returns $P=0.988$, meaning that the restricted BAT and main samples follow the same distribution in $\sigma$ (with almost 99\,\% of probability), while including the complementary sample, this probability decreases to 77\,\%, being still high. 

Considering that the central $\sigma$ values are representative of the bulge of the galaxies, we can use the $M_\bullet-\sigma$ relation \citep[e.g.][]{ferrarese00,gebhardt00,tremaine02,ferrarese05,graham11} to determine the mass of the central supermassive black hole ($M_\bullet$).  Using equation~3 from \citet{kormendy13} and the $\sigma$ values from Fig.~\ref{hkin}, we obtain $(0.15 \lesssim M_\bullet \lesssim 13.5)\times10^8\,$M$_\odot$  and  mean values of $<M_\bullet>\approx1.3\times10^{8}$\,M$_\odot$ for the main sample and  $<M_\bullet>\approx9.8\times10^{7}$\,M$_\odot$ including the complementary sample.

The main goals of our project are to map and quantify AGN feeding and feedback process via gas inflows and outflows. While inflows are usually restricted to the plane of the galaxy disk \citep[e.g.][]{n4051,mrk79}, outflows do not show any preferential orientation \citep[][]{schmitt01,barbosa14,n5548}. Thus, in order to optimize the search for inflows and outflows, it is desirable that the sample of galaxies show a wide range of disk orientations. The bottom panels of Figure~\ref{hkin} show histograms for the axial ratio $b/a$ for our sample and restricted BAT sample (where $a$ and $b$ are the semi-major and semi-minor axes of the galaxy obtained from the Hyperleda database, measured at the isophote 25 mag/arcsec$^2$ in the  B-band surface brightness distribution). Measurements of the axial ratio are available for all galaxies of our sample and for 59 objects of the restricted BAT sample. The bottom panels of Fig.~\ref{hkin} show histograms for the axial ratio in bins of 0.05.  The mean values of $b/a$ of our main samples are similar to that of the restricted BAT sample and including the complementary sample.  Our sample shows a wide range of axial ratios, from nearly edge-on galaxies ($b/a\sim0.2$, corresponding to a disk inclination $i\sim80^\circ$) to nearly face-on galaxies ($b/a\sim0.9$, $i\sim25^\circ$). The K-S test shows a probability of 88\,\% of the main and restricted BAT samples follow the same distribution in axial ratio, while including the complementary sample, the K-S test results in $P=0.416$, suggesting that the complementary sample includes a bias in the axial ratio distribution. 


\subsection{Observations}\label{observations}

The Integral Field Spectroscopic observations of the galaxies of our sample have been obtained with the Gemini Near-Infrared Integral Field Spectrograph \citep[NIFS,][]{mcgregor03} operating with the Gemini North adaptive optics module ALTAIR. NIFS has a square field of view of $\approx3\farcs0\times3\farcs0$, divided into 29 slices with an angular sampling of 0$\farcs$1$\times$0$\farcs$04. The observations of our sample are part of a Large and Long Program (LLP) approved by Brazilian National Time Allocation Committee (NTAC) and have started in semester 2015A and are planned to be concluded in 2019B. Some galaxies shown in Table~\ref{sampletab} were observed as part of previous proposals by our group. The data comprise J and K(K$_l$)-band observations at angular resolutions in the range 0\farcs12--0\farcs20, depending on the performance of the adaptive optics module and velocity resolution of about 40\,\kms\ at both bands. 

Emission lines from high, low-ionization and molecular gas, as well as strong CO absorptions,  are usually observed at these spectral bands in spectra of active galaxies \citep[e.g.][]{rogerio06}, allowing the mapping of the gas kinematics, distribution, excitation, extinction and the stellar kinematics. The relatively high spatial and spectral resolutions, together with the spatial coverage, make this an unprecedented data set to map the AGN feeding and feedback processes in nearby galaxies.  The on-source exposure time for each galaxy is in the range 0.7--1.7 hours at each band, expected to result in a signal-to-noise ratio $snr>10$, which allows the fitting of the emission and absorption lines. The observations have been following the standard object-sky-object dithering sequence and the data reduction have been done following the standard procedures of spectroscopic data treatment. 

\subsection{Data reduction}

The data reduction for the J and K band are being performed following the same procedure used in previous works \citep[e.g.][]{n4051,diniz15,stel_llp}, including the trimming of the images, flat-fielding, sky subtraction, wavelength and s-distortion calibrations and correction of the telluric absorptions. The spectra are then flux calibrated  by interpolating a black body function to the spectrum of the telluric standard star. Finally, datacubes for each individual exposure are created with an angular sampling of 0\farcs05$\times$0\farcs05. These cubes are then mosaicked using the continuum peak as reference and median combined to produce a single final datacube for each band.

\begin{table*} 
\centering
\caption{Observations. (1) Galaxy name;  (2) Gemini project identification; (3) J and (4) K-band on-source exposure time; (5) J and (6) K-band angular resolution estimated from the FWHM of the flux distribution of the telluric standard star; (7) J and (8) K-band spectral resolution estimated from the FWHM of the Arc Lamp lines used for wavelength calibrate the datacubes; (9) References to published studies using this dataset.
}
\vspace{0.3cm}
\begin{tabular}{l c c c c c c c l }
\hline
       (1)         &   (2)     &    (3)            &  (4)   &    (5)    & (6)	     & (7) & (8)  & (9) \\
Galaxy 	           &Programme      & J Exp. T.  & K Exp. T.  & PSF$_{\rm J}$   &PSF$_{\rm K}$ &FWHM$_{\rm J}$ &FWHM$_{\rm K}$ & Refs.	 \\
                   &               & (seconds)  &  (seconds) &    (arcsec)        & (arcsec)  & (\AA)         & (\AA)         &   \\
\hline								    		   		       
\multicolumn{7}{c}{Main sample}\\
\hline 
         NGC788  &  GN-2015B-Q-29 & 7$\times$400 & 11$\times$400 & 0.13	&  0.13    & 1.9    & 3.5  & a \\
        NGC1068  &GN-2006B-C-9$^*$& 27$\times$90 & 27$\times$90  & 0.14	& 0.11 	   & 1.7     & 3.0  & b, c, d\\
        NGC2110  &  GN-2015B-Q-29 & 6$\times$400 &  --           &  0.13&   --	   & 1.9& -- & a, e \\
            --   & GN-2010B-Q-25  &     -        & 6$\times$600  & 0.15	& --       & -- & 3.4&  --\\
           Mrk3  &GN-2010A-Q-5$^*$& 6$\times$600 &  6$\times$600 &0.13  &  0.13	   & 2.0&3.2 & -- \\
        NGC3227  & GN-2016A-Q-6   & 6$\times$400 & 6$\times$400  & 0.13	&  0.12	   &1.8 & 3.5& a	\\
        NGC3516  & GN-2015A-Q-3   & 10$\times$450& 10$\times$450 & 0.17	&   0.15   & 1.8& 3.5& a	    \\
        NGC4151  &GN-2006B-C-9$^*$& 8$\times$90  &8$\times$90    &0.16	& 0.12 	   &1.6 &3.3 & f, g, h  \\
        NGC4235  &  GN-2016A-Q-6  & 9$\times$400 & 10$\times$400 & 0.12	&  0.13	   & 1.8& 3.5& a	   \\
         Mrk766  &  GN-2010A-Q-42 & 6$\times$550 & 6$\times$550  &0.21 	& 0.19	   & 1.7& 3.5& a, j \\
        NGC4388  & GN-2015A-Q-3   &  --          & 2$\times$400  &  --	&  0.19	   & -- & 3.7& a	    \\
        NGC5506  &  GN-2015A-Q-3  &10$\times$400 & 10$\times$400 & 0.15	&   0.18   &1.9 & 3.6& a		\\
\hline
\multicolumn{7}{c}{Complementary Sample} \\
\hline  															        
NGC1052   & GN-2010B-Q-25    & 6$\times$610 & 4$\times$600  & 	 & 0.15	  &   1.7 & & a \\
NGC4051   & GN-2006A-SV-123  & --           & 6$\times$750  & --	 & 0.18	  & --    & 3.2 & a, k \\
NGC5548   & GN-2012A-Q-57    & 12$\times$450& 12$\times$450 & 0.28	 & 0.20	  &  1.7  & 3.5 & a, l \\
NGC5899   & GN-2013A-Q-48    & 10$\times$460& 10$\times$460 & 0.13	 & 0.13	  &  1.8  & 3.4 & a	 \\
NGC5929   & GN-2011A-Q-43    & 10$\times$600& 10$\times$600 & 0.12	 & 0.12	  &  1.7  & 3.2  & a, m, n	  \\
Mrk79     & GN-2010A-Q-42    & 6$\times$520 & 6$\times$550  & 0.25	 & 0.25	  &  1.8  & 3.5  & o \\
Mrk607    &  GN-2012B-Q-45   &10$\times$500 & 12$\times$500 & 0.14	 & 0.14	  &    2.0 & 2.2 & a	  \\
Mrk1066   &  GN-2008B-Q-30   & 8$\times$600 & 8$\times$600  & 0.13	 & 0.15	  &  1.7  & 3.3 & a, p, q, r, s \\
Mrk1157   &  GN-2009B-Q-27   & 8$\times$550 & 8$\times$550  & 0.11	 & 0.12	  &  1.8  & 3.5 & a, t, u \\
\hline									
\multicolumn{9}{l}{$^*$ From Gemini Science Archive}\\
\multicolumn{9}{l}{References: a: \citet{stel_llp}; b: \citet{sb12}; c: \citet{n1068-exc}; d: \citet{barbosa14}; }\\
\multicolumn{9}{l}{ e: \citet{diniz15}; f: \citet{sb09}, g: \citet{sb10};   }\\
\multicolumn{9}{l}{h: \citet{n4151-torus}; i: \citet{mrk766}; k: \citet{n4051}; l: \citet{n5548};  }\\
\multicolumn{9}{l}{m: \citet{n5929-let}; n: \citet{n5929};  }\\
\multicolumn{9}{l}{o: \citet{mrk79}; p: \citet{mrk1066-pop}; q: \citet{mrk1066-exc}; }\\
\multicolumn{9}{l}{r: \citet{mrk1066-kin};  s: \citet{ra14}; t: \citet{mrk1157-pop}; }\\
\multicolumn{9}{l}{u: \citet{mrk1157}.  }\\
\hline									
\end{tabular}
  \label{sampleobs}
\end{table*}

Table~\ref{sampleobs} presents a summary of the observation logs for the galaxies already observed. The angular resolution at J (PSF$_J$) and K (PSF$_K$) was estimated by measuring the full width at half maximum (FWHM) of the telluric standard star flux distributions. The uncertainties in the measurements are about 0\farcs03 for all galaxies at both bands. The spectral resolution at the J and K band was estimated from the FWHM of emission lines of the Ar and ArXe lamps used to wavelength calibration, respectively. For the J band we fitted the profiles of typical lines observed near 1.25~$\mu$m, while for the K band the spectral resolution was estimated from lines seen around 2.2~$\mu$m.  The spectral resolution ranges from 1.7 to 2.0\,\AA\ at the J band, corresponding to an instrumental broadening ($\sigma_{inst}=\frac{FWHM}{2.355}\frac{c}{\lambda_c}$) of 17--20~\kms. At the K band the spectral resolutions ranges from 3 to 3.7~\AA, translating into $\sigma_{inst}\approx17-21$\,\kms.


\section{Molecular and ionized gas surface mass density} \label{massdens}

We use the available data to discuss the radial distribution of ionized and molecular gas for galaxies already observed. The fluxes of the H$_2\lambda$2.12\,$\mu$m and Br$\gamma$ emission lines can be used to estimate the mass of hot molecular and ionized gas, respectively. Following \citet{osterbrock06} and \citet{sb09}, the mass of ionized  ($M_{\rm H\,II}$)  gas can be obtained from

\begin{equation}
 \left(\frac{M_{\rm H\,II}}{M_\odot}\right)=3\times10^{19}\left(\frac{F_{\rm Br\gamma}}{\rm erg\,cm^{-2}\,s^{-1}}\right)\left(\frac{D}{\rm Mpc}\right)^2\left(\frac{N_e}{\rm cm^{-3}}\right)^{-1},
\label{mhii}
\end{equation}
where $D$ is the distance to the galaxy, $F_{\rm Br\gamma}$ is the Br$\gamma$ flux and $N_e$ is the electron density,  assuming an electron temperature of 10$^4$K. We have adopted an electron density of $N_e=500$\,cm$^{-3}$, which is a typical value for the inner few hundred pcs of AGNs as determined from the [S\,{\sc ii}]$\lambda\lambda$6717,6730  lines \citep[e.g.][]{dors14,brum17}.

Under the assumptions of local thermal equilibrium and excitation temperature of 2000\,K, the mass of hot molecular  gas ($M_{\rm H2}$) can be obtained from \citep[e.g.][]{scoville82,n1068-exc}:
 
\begin{equation}
 \left(\frac{M_{H_2}}{M_\odot}\right)=5.0776\times10^{13}\left(\frac{F_{H_{2}\lambda2.1218}}{\rm erg\,s^{-1}\,cm^{-2}}\right)\left(\frac{D}{\rm Mpc}\right)^2,
\label{mh2}
\end{equation}
where $F_{H_{2}\lambda2.1218}$ is the H$_2$ (2.1218$\mu$m) emission-line flux.

\begin{figure*}
\includegraphics[width=0.48\textwidth]{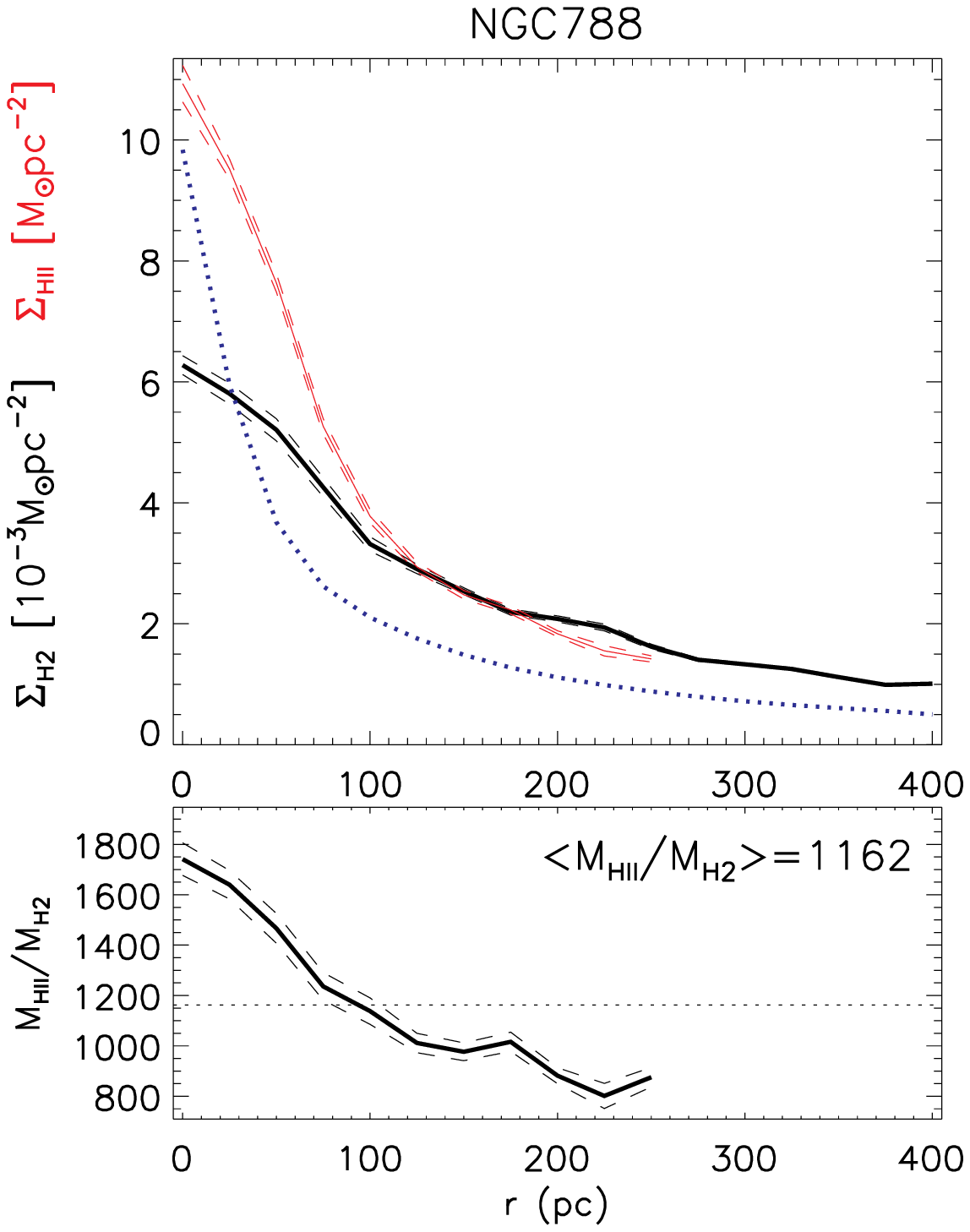}
\includegraphics[width=0.48\textwidth]{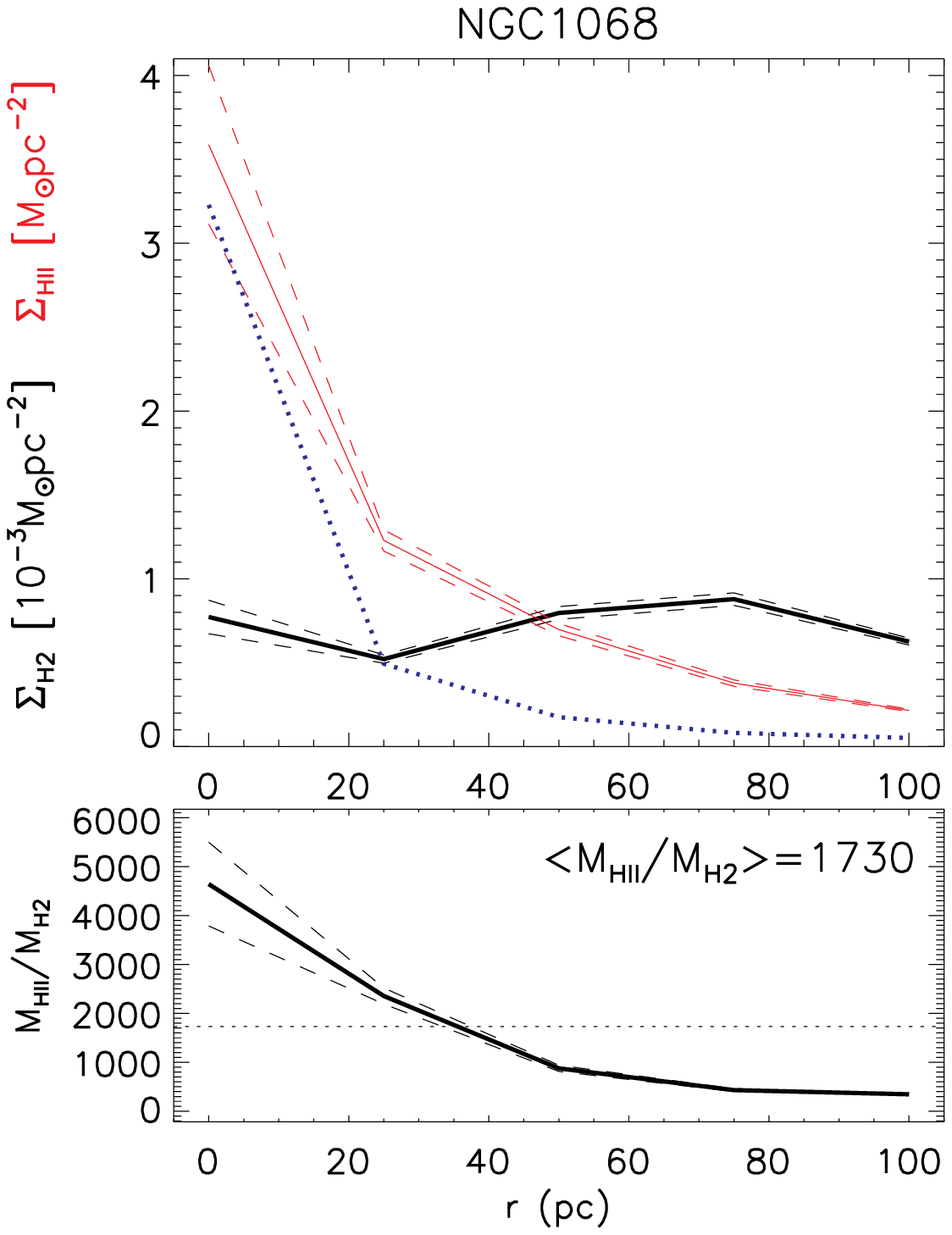}
\includegraphics[width=0.48\textwidth]{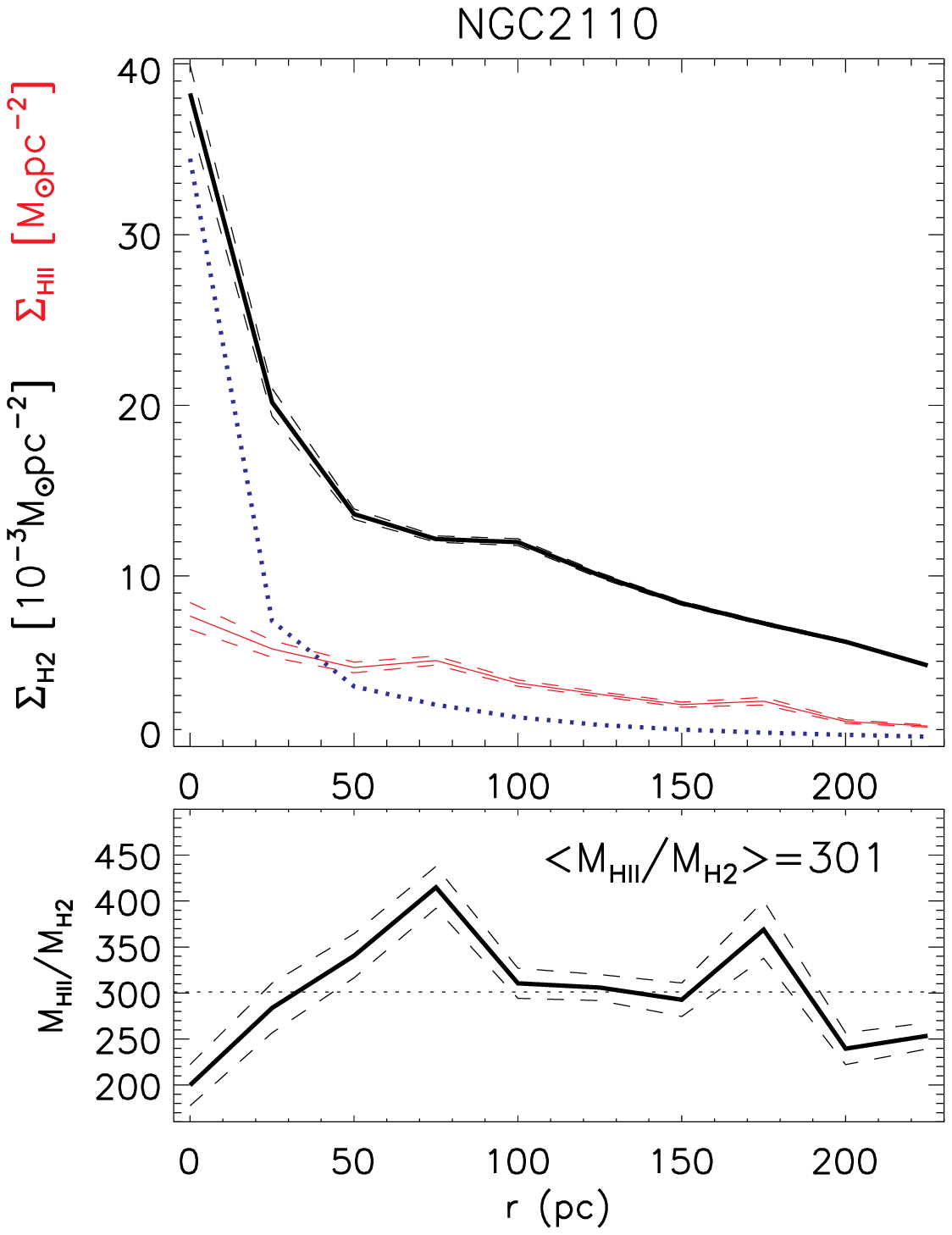}
\includegraphics[width=0.48\textwidth]{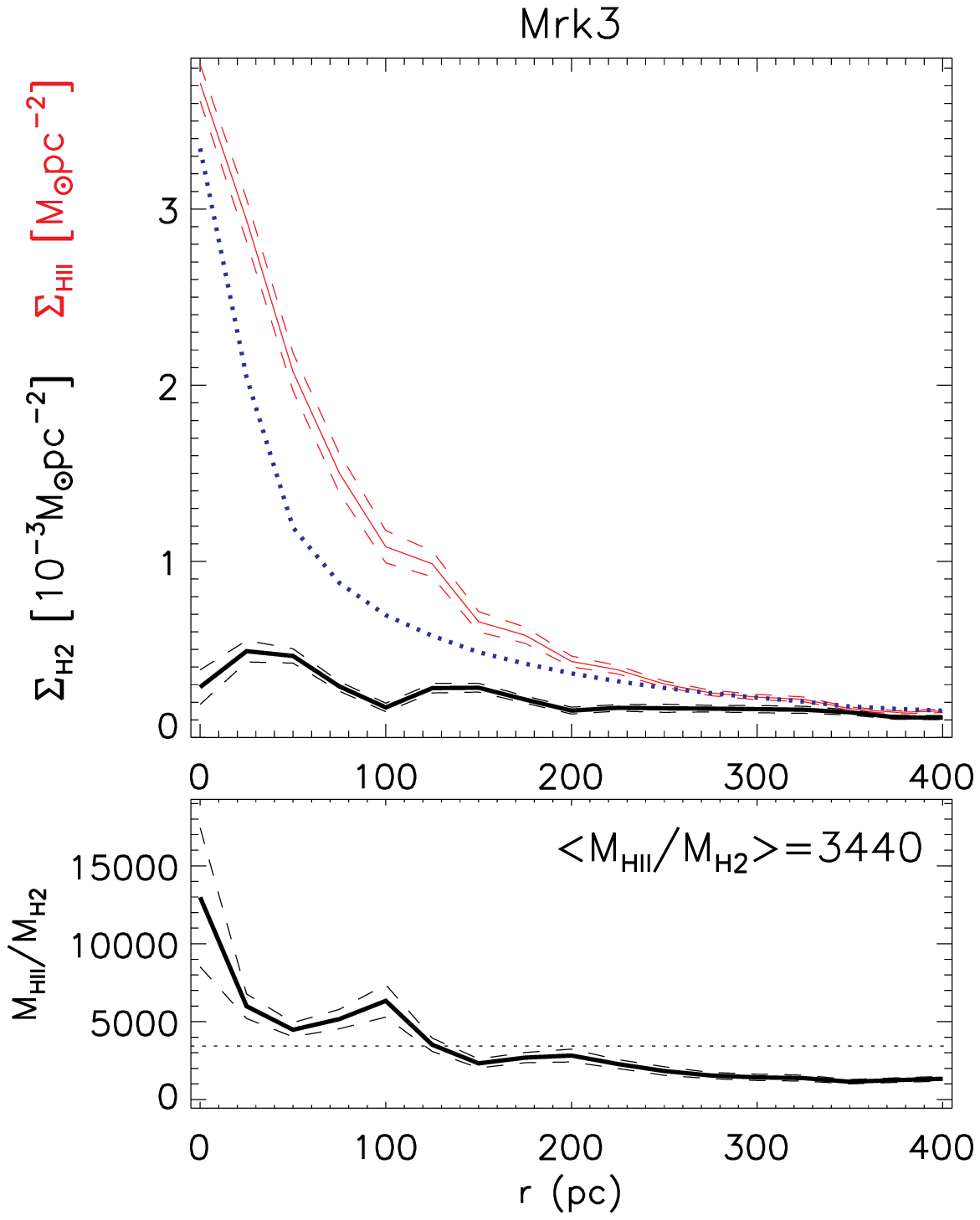}
\caption{The top panels show the surface mass density profiles for the hot molecular (black) and ionized (red) for a radial bin of 25~pc at the plane of the galaxy. The profiles are shown as continuous lines and the dashed lines shows the standard error variation. The K-band surface brightness is shown as a dotted blue line in units of $C\times$erg s$^{-1}$ cm$^{-2}$ \AA$^{-1}$ arcsec$^{-2}$, where $C$ is an arbitrary constant. The bottom panels show the ratio ratio between the mass of ionized and molecular gas variation for the same radial bin, considering only spaxels with measurements of both masses. The mean value of the ratio $<M_{\rm HII}/M_{\rm H2}>$ is shown at the top-right corner of the corresponding panel. The geometric parameters of the disk, used in the deprojection are shown in Table\,\ref{gasdens}. In this figure, we show the profiles for NGC\,788, NGC\,1068, NGC\,2110 and Mrk\,3.}
\label{sdenp}
\end{figure*}

\begin{figure*}
\includegraphics[width=0.48\textwidth]{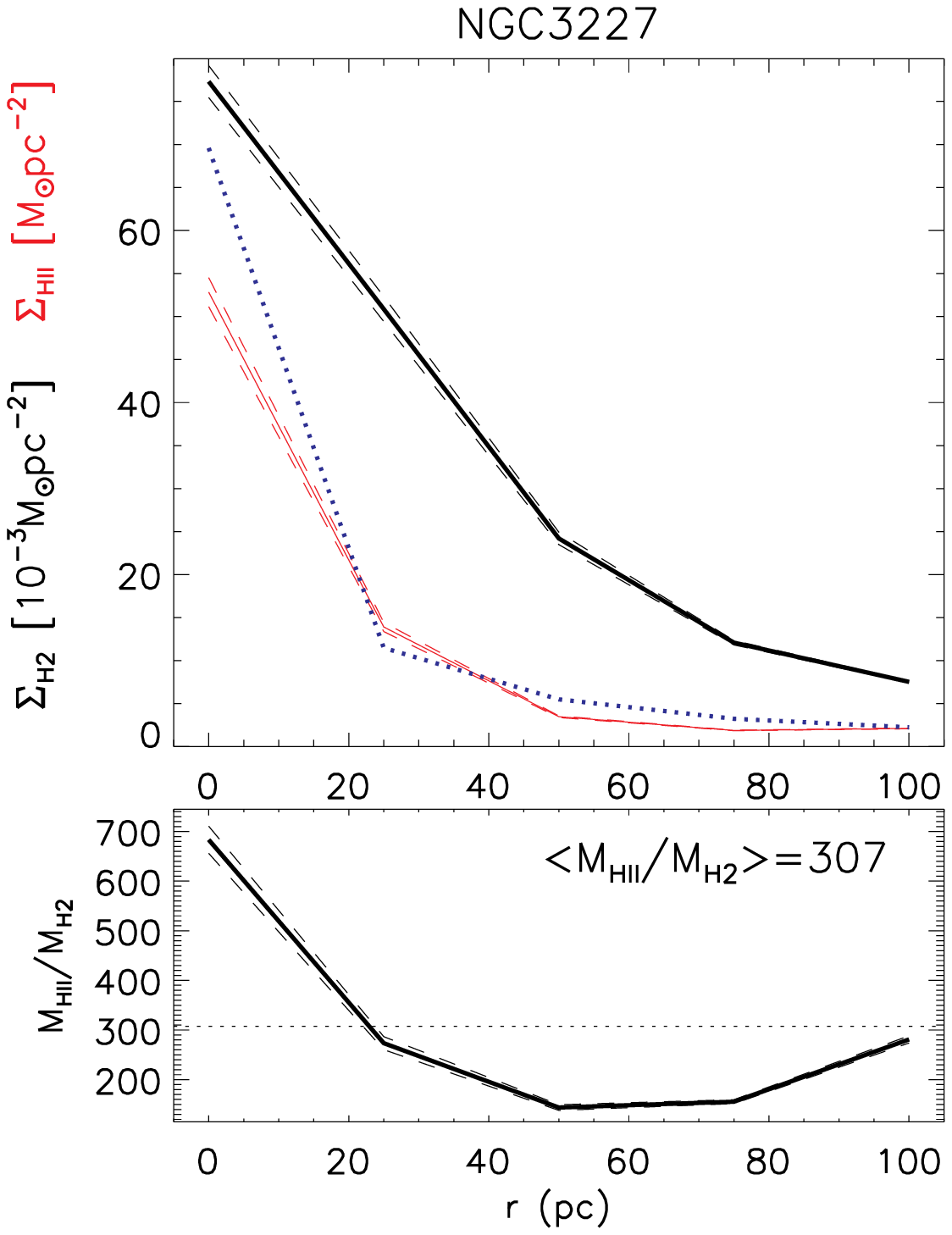}
\includegraphics[width=0.48\textwidth]{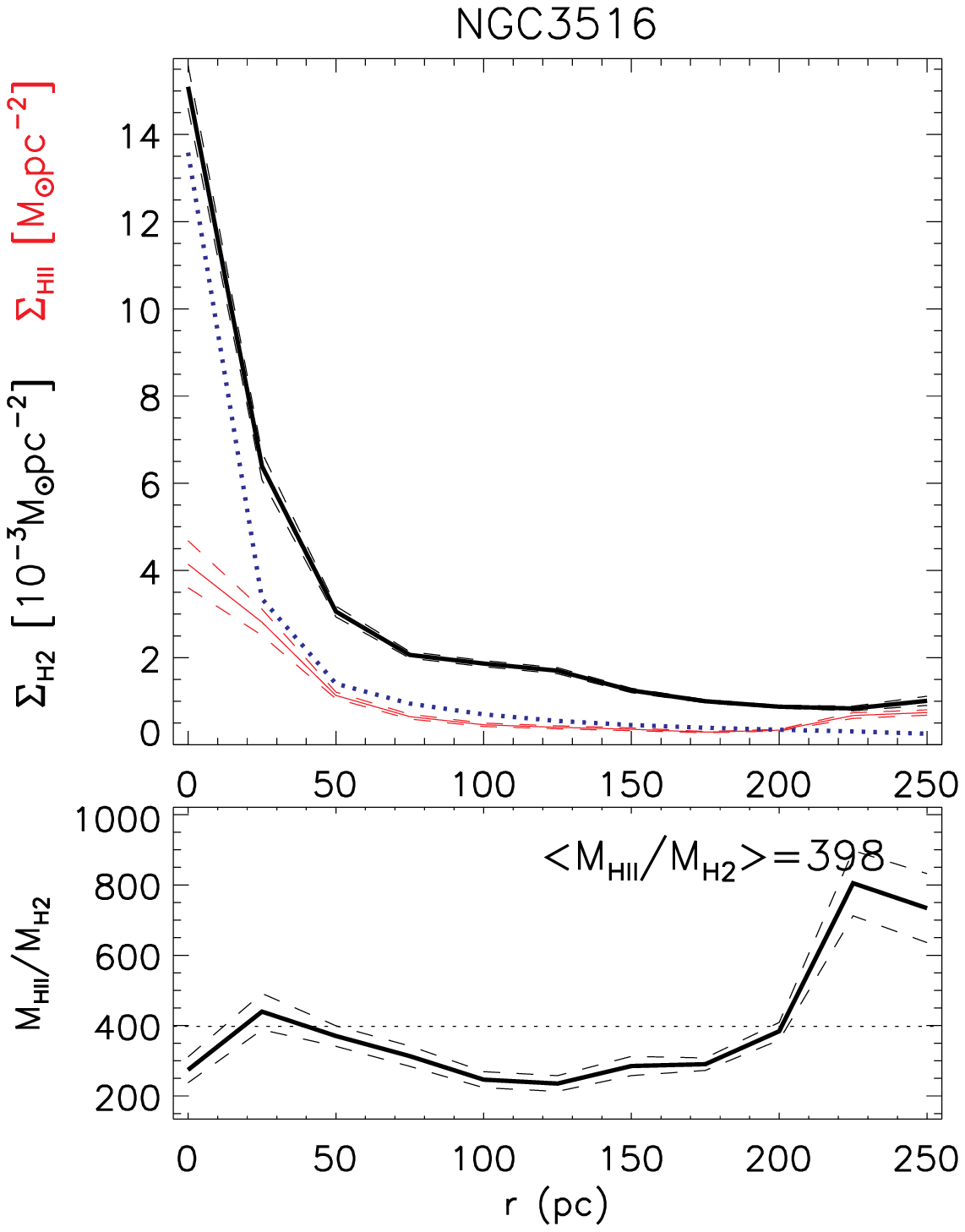}
\includegraphics[width=0.48\textwidth]{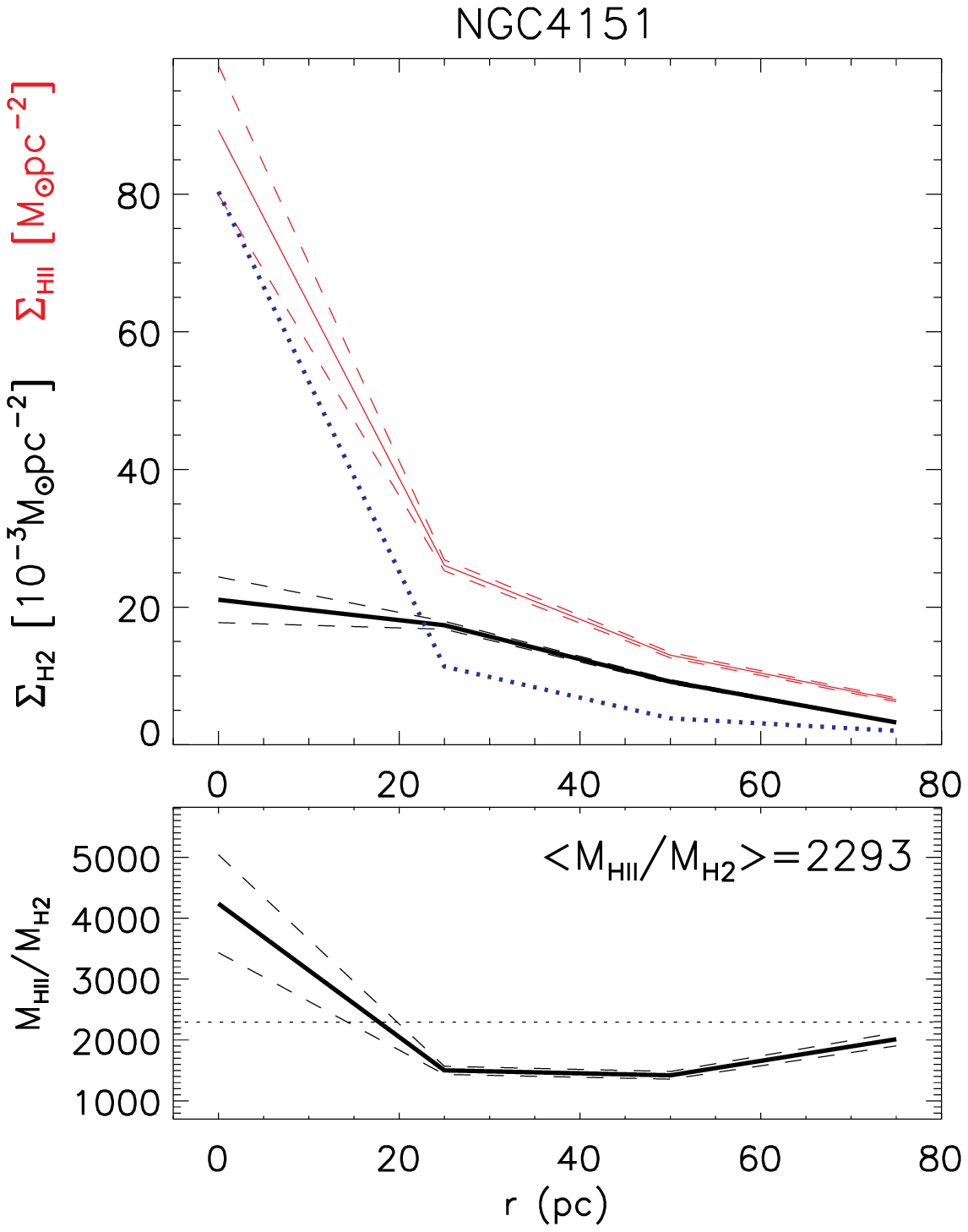}
\includegraphics[width=0.48\textwidth]{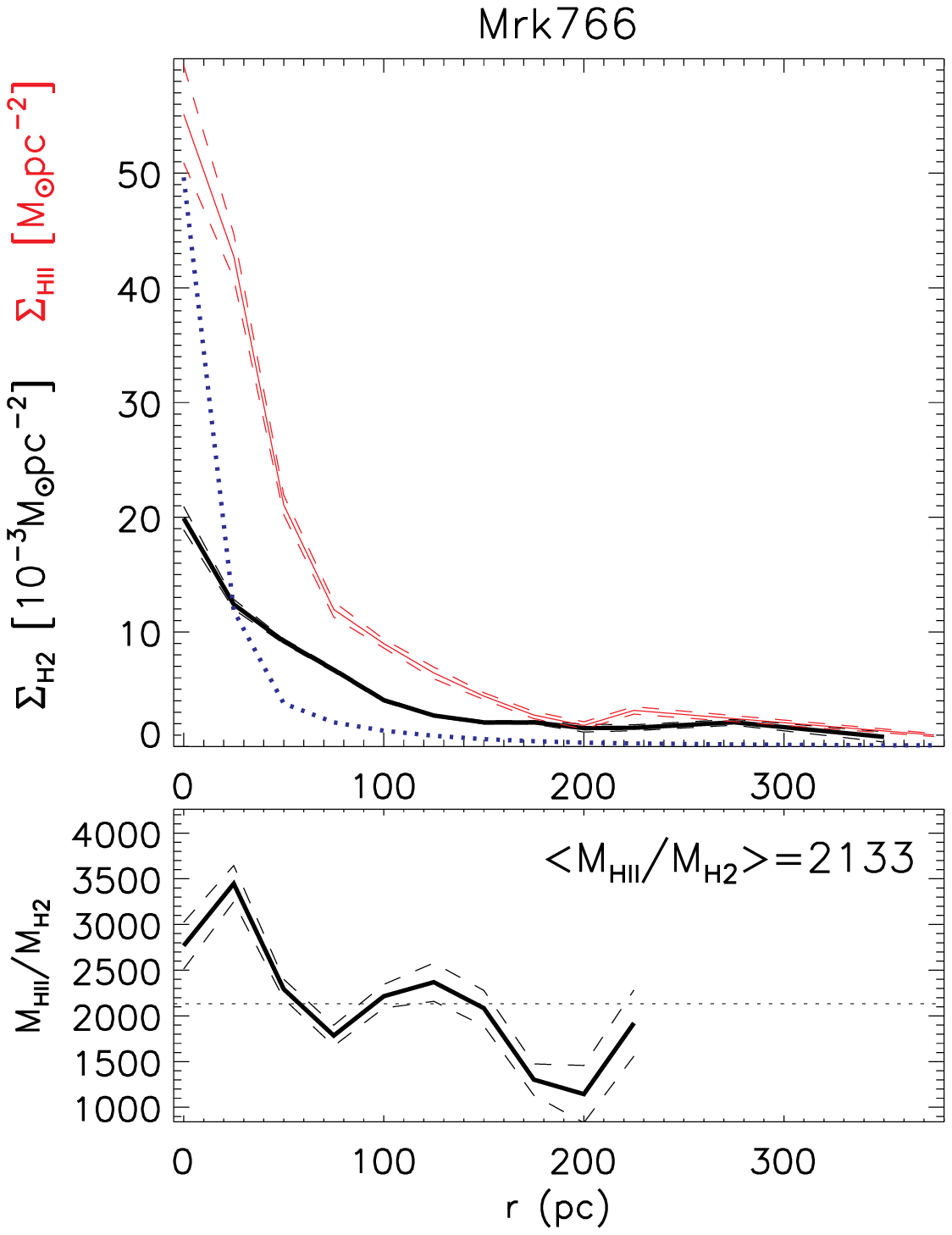}
\caption{Same as Fig.~\ref{sdenp} for NGC\,3227, NGC\,3516, NGC\,4151 and Mrk\,766.}
\label{sdenpa}
\end{figure*}

\begin{figure*}
\includegraphics[width=0.48\textwidth]{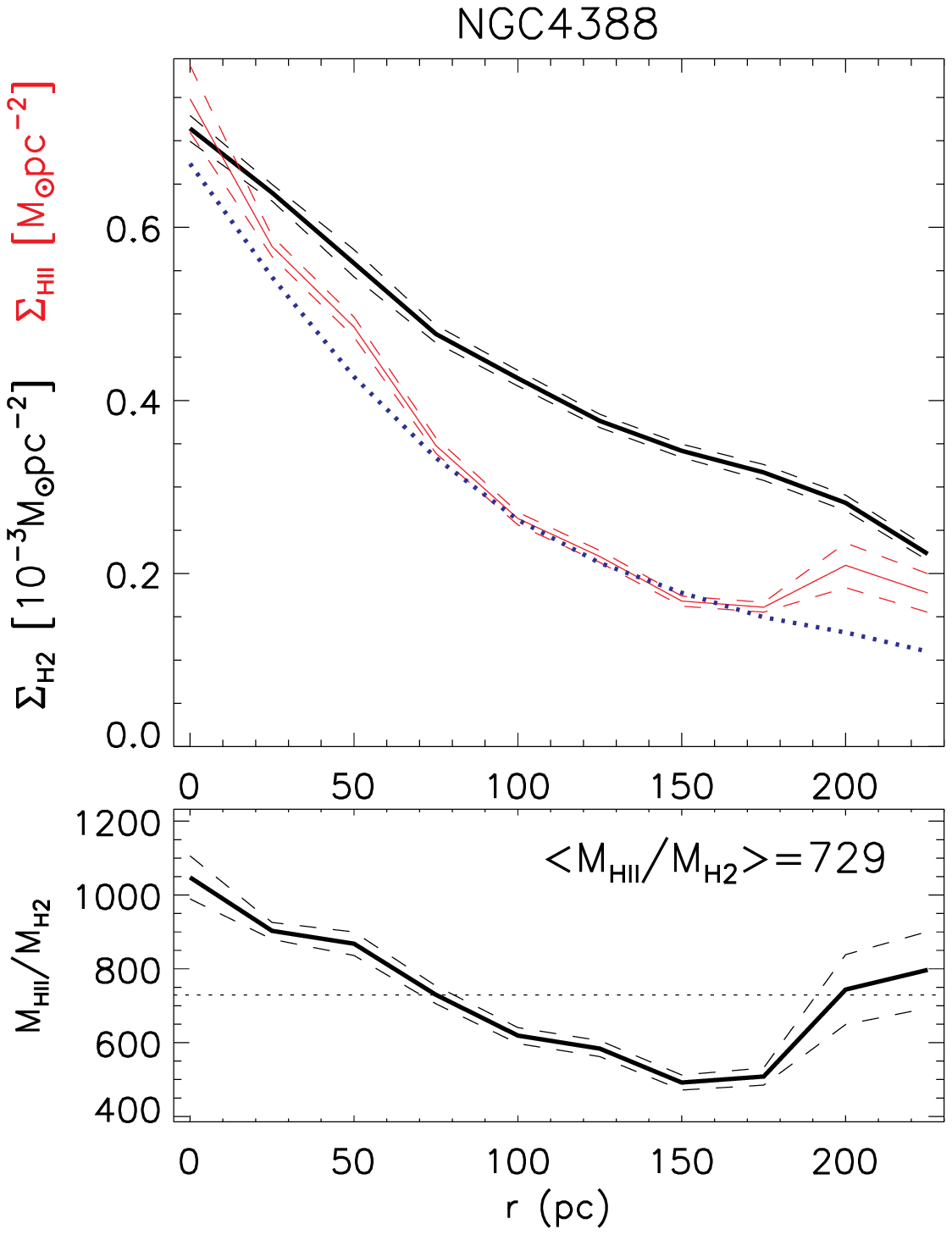}
\includegraphics[width=0.48\textwidth]{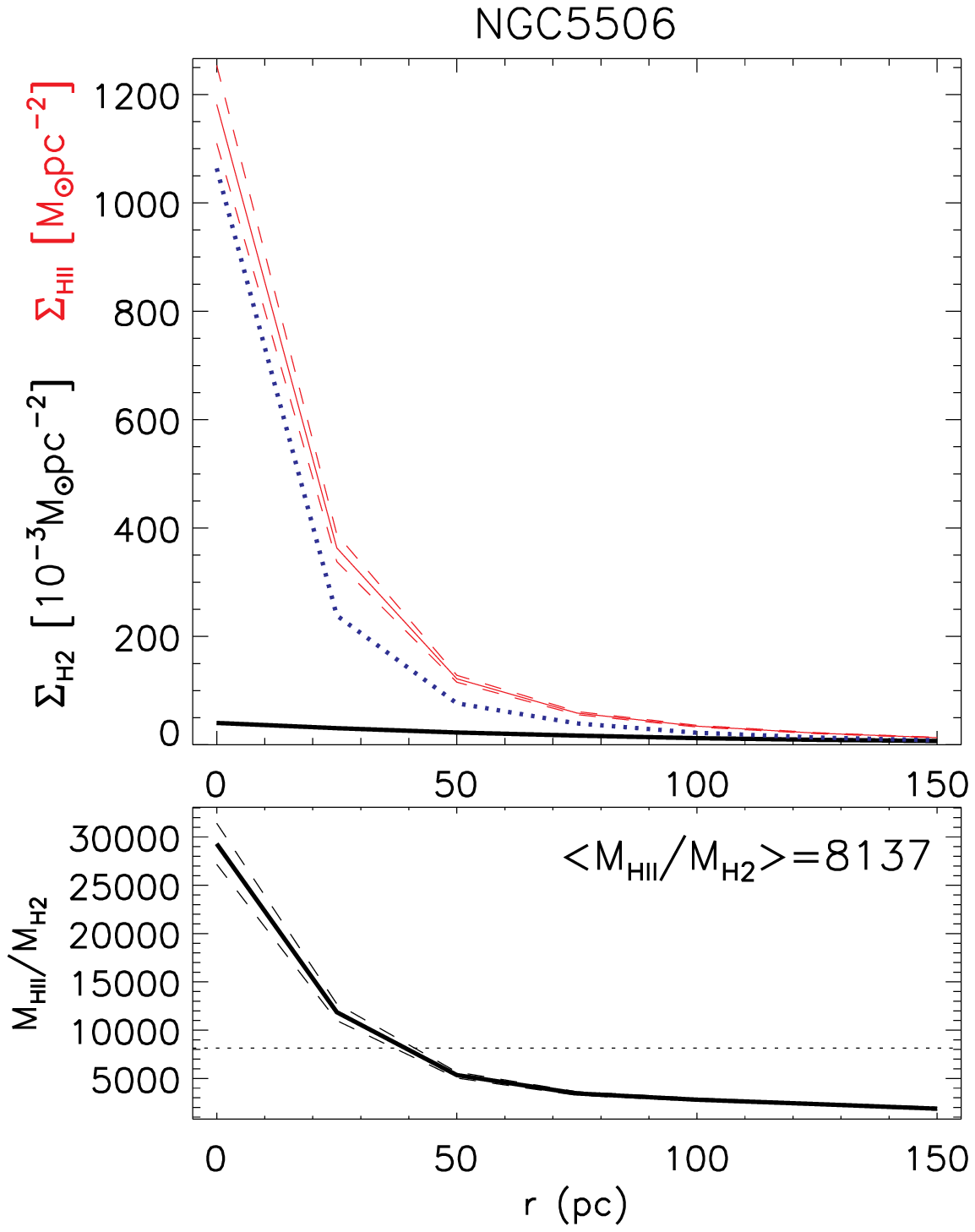}
\includegraphics[width=0.48\textwidth]{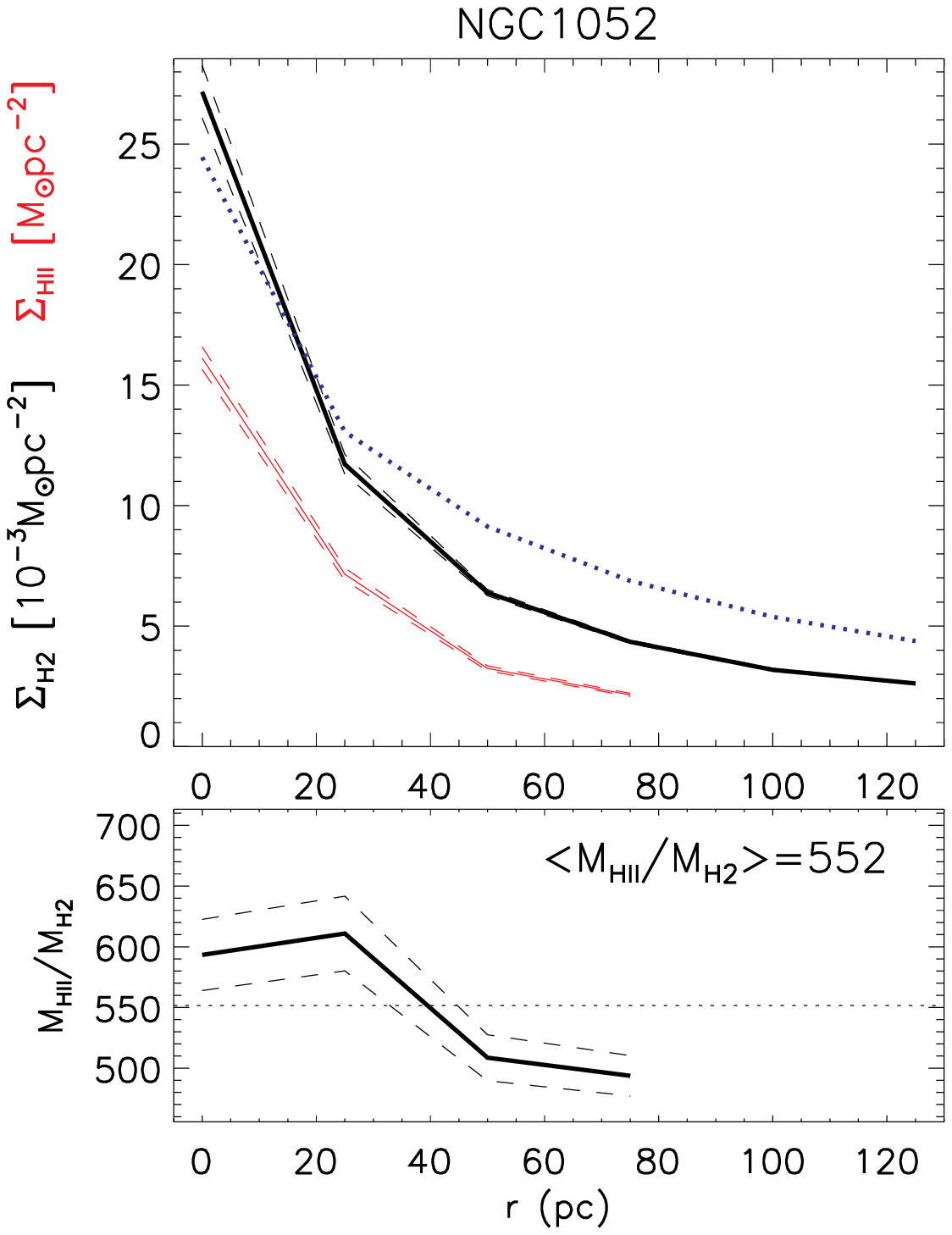}
\includegraphics[width=0.48\textwidth]{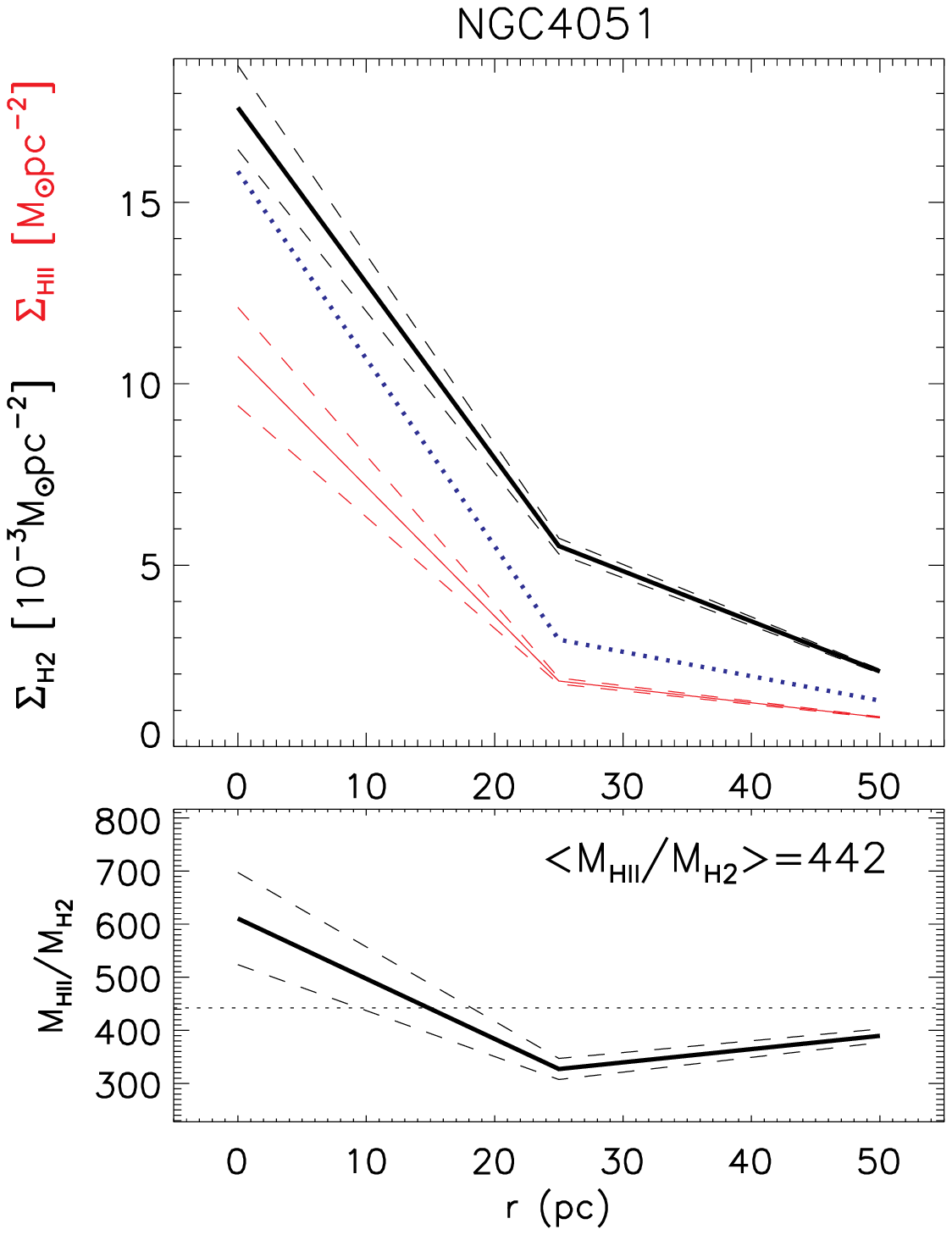}
\caption{Same as Fig.~\ref{sdenp} for NGC\,4388, NGC\,5506, NGC\,1052 and NGC\,4051.}
\label{sdenpb}
\end{figure*}

\begin{figure*}
\includegraphics[width=0.48\textwidth]{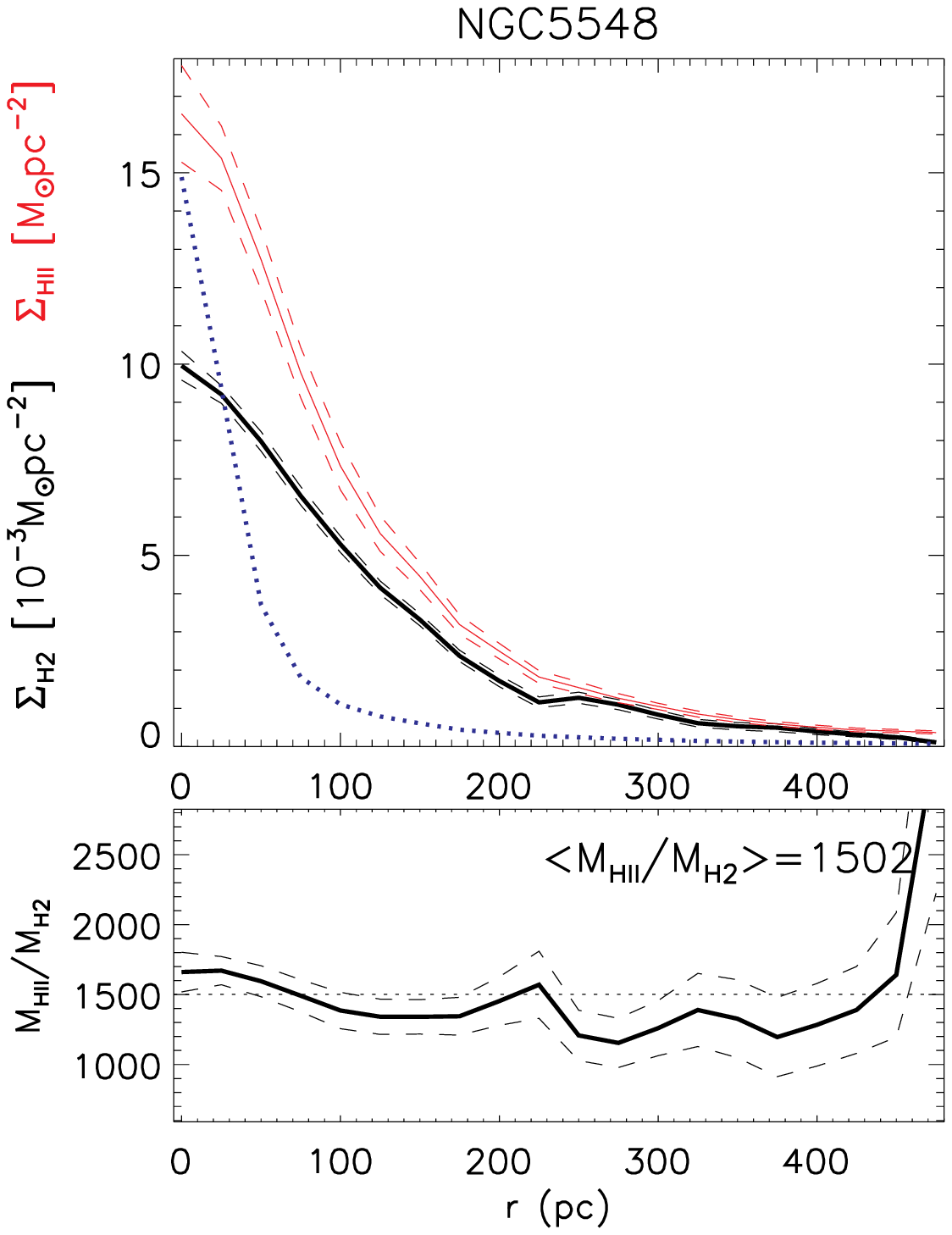}
\includegraphics[width=0.48\textwidth]{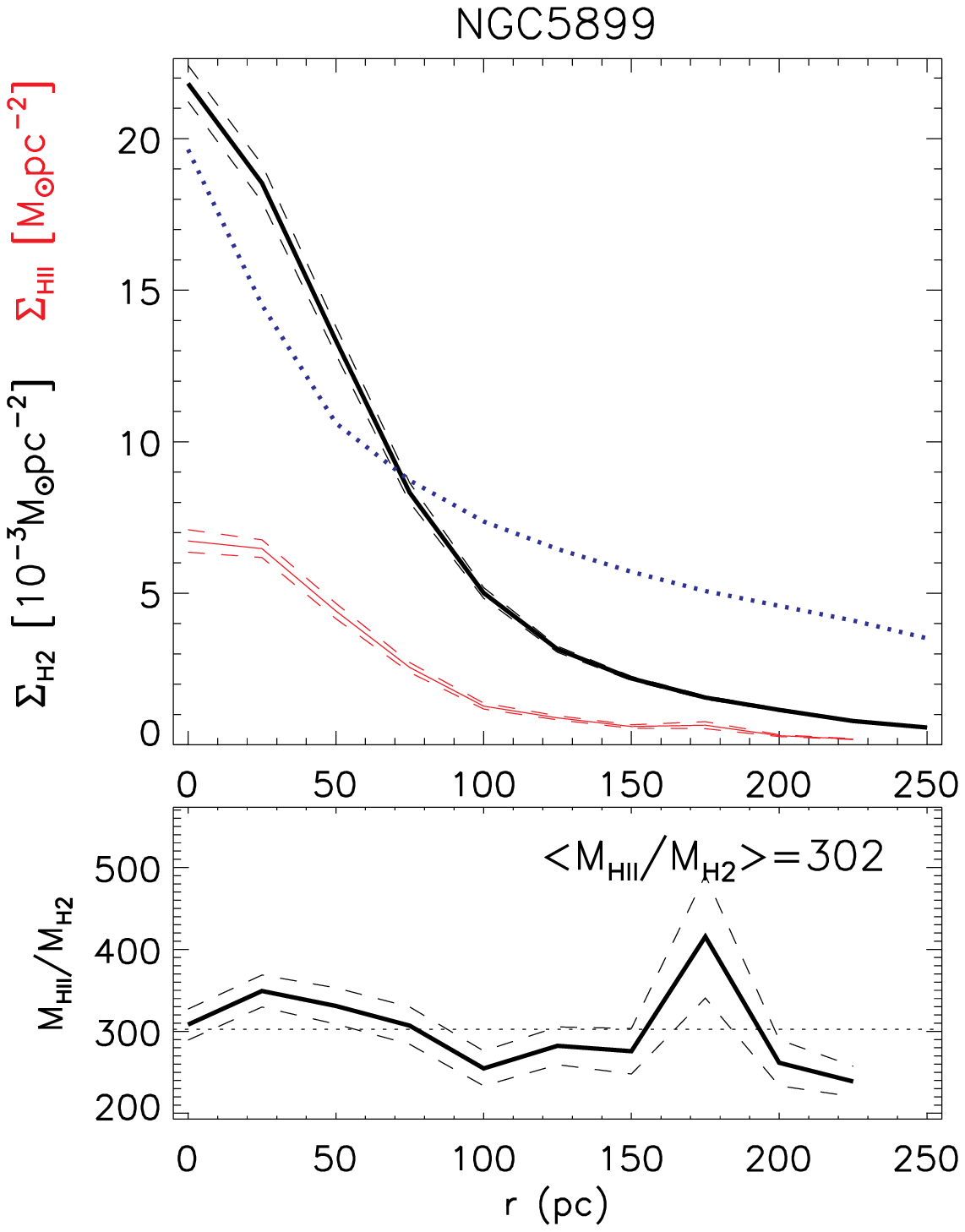}
\includegraphics[width=0.48\textwidth]{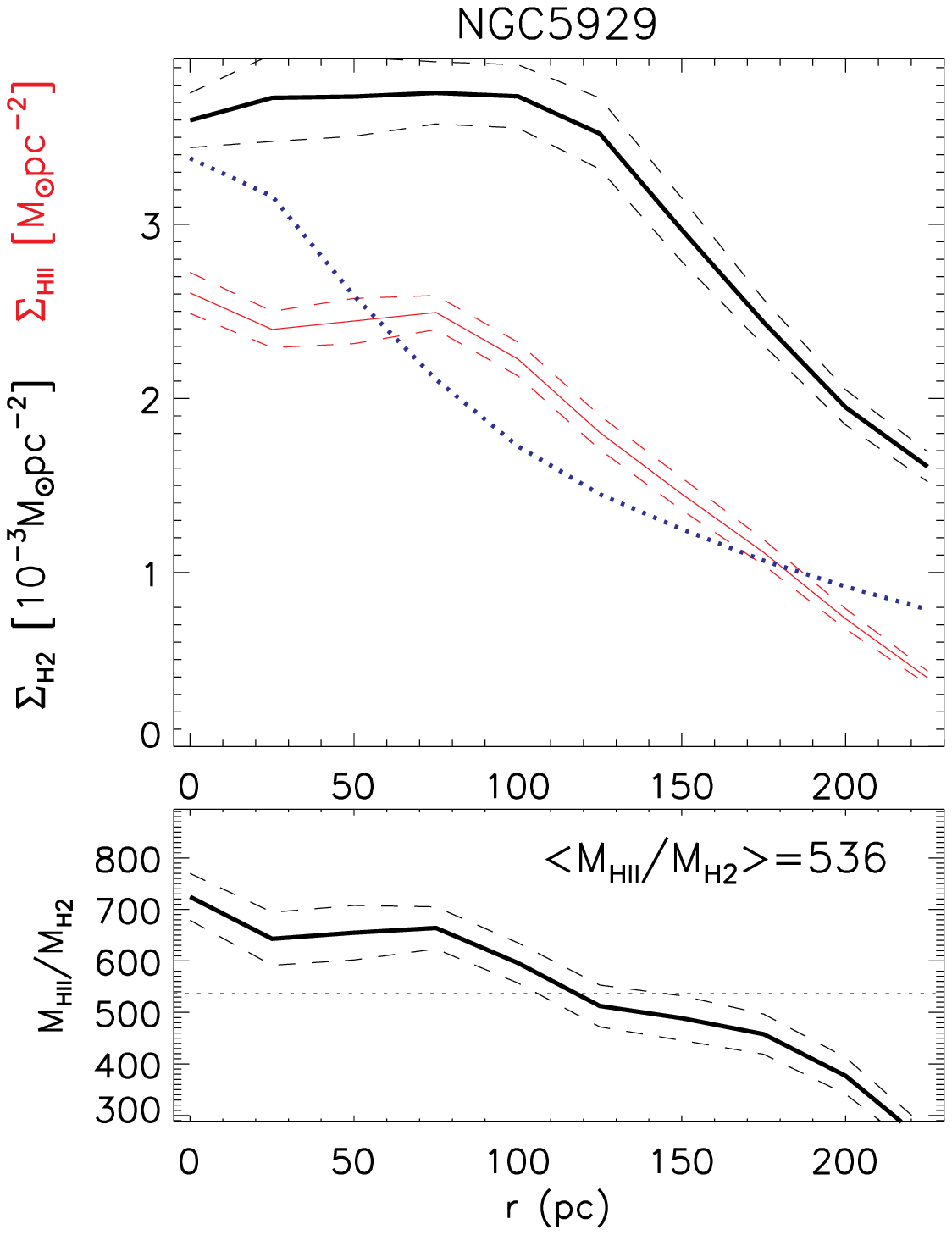}
\includegraphics[width=0.48\textwidth]{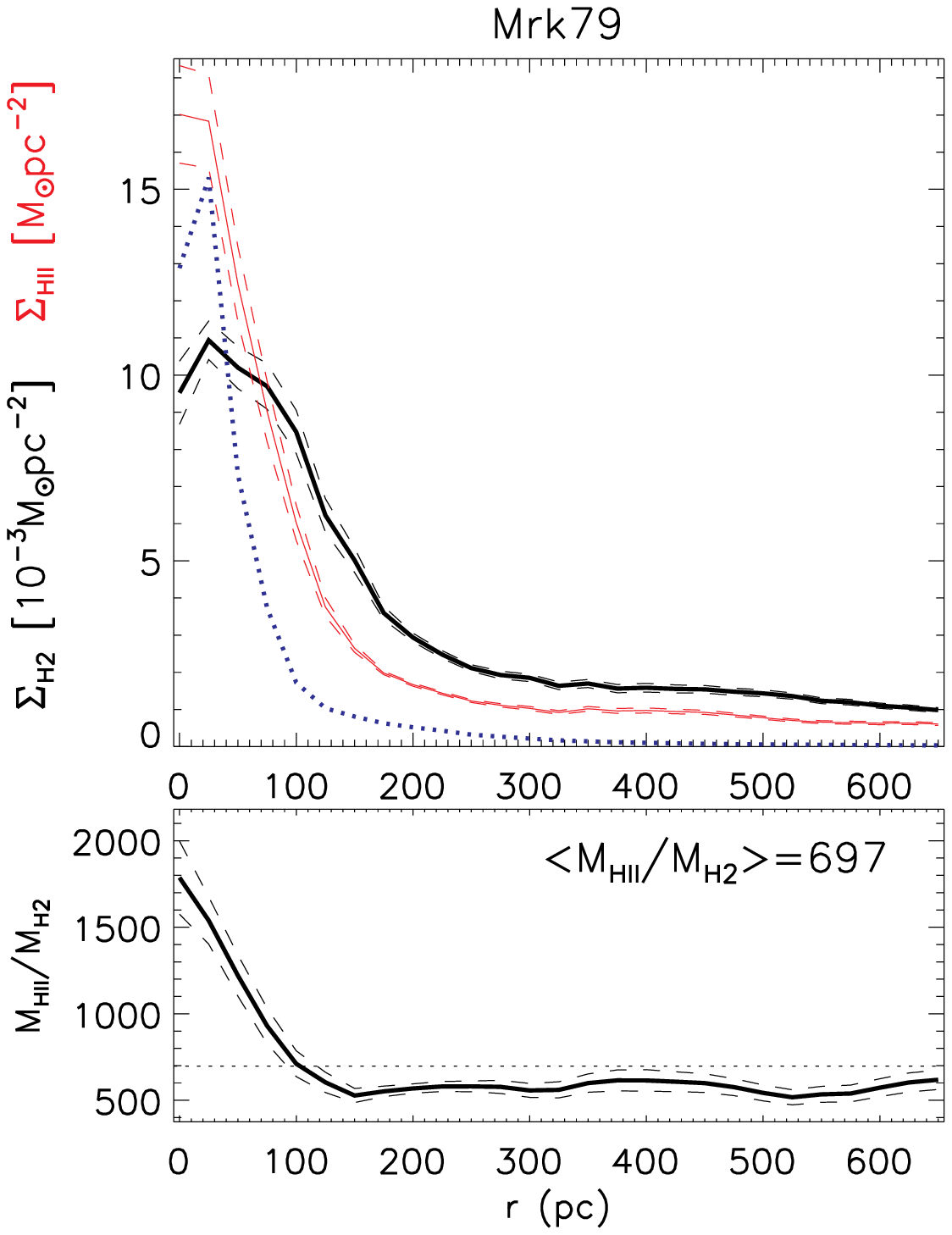}
\caption{Same as Fig.~\ref{sdenp} for NGC\,5548, NGC\,5899, NGC\,5929 and Mrk\,79.}
\label{sdenpc}
\end{figure*}

\begin{figure*}
\includegraphics[width=0.48\textwidth]{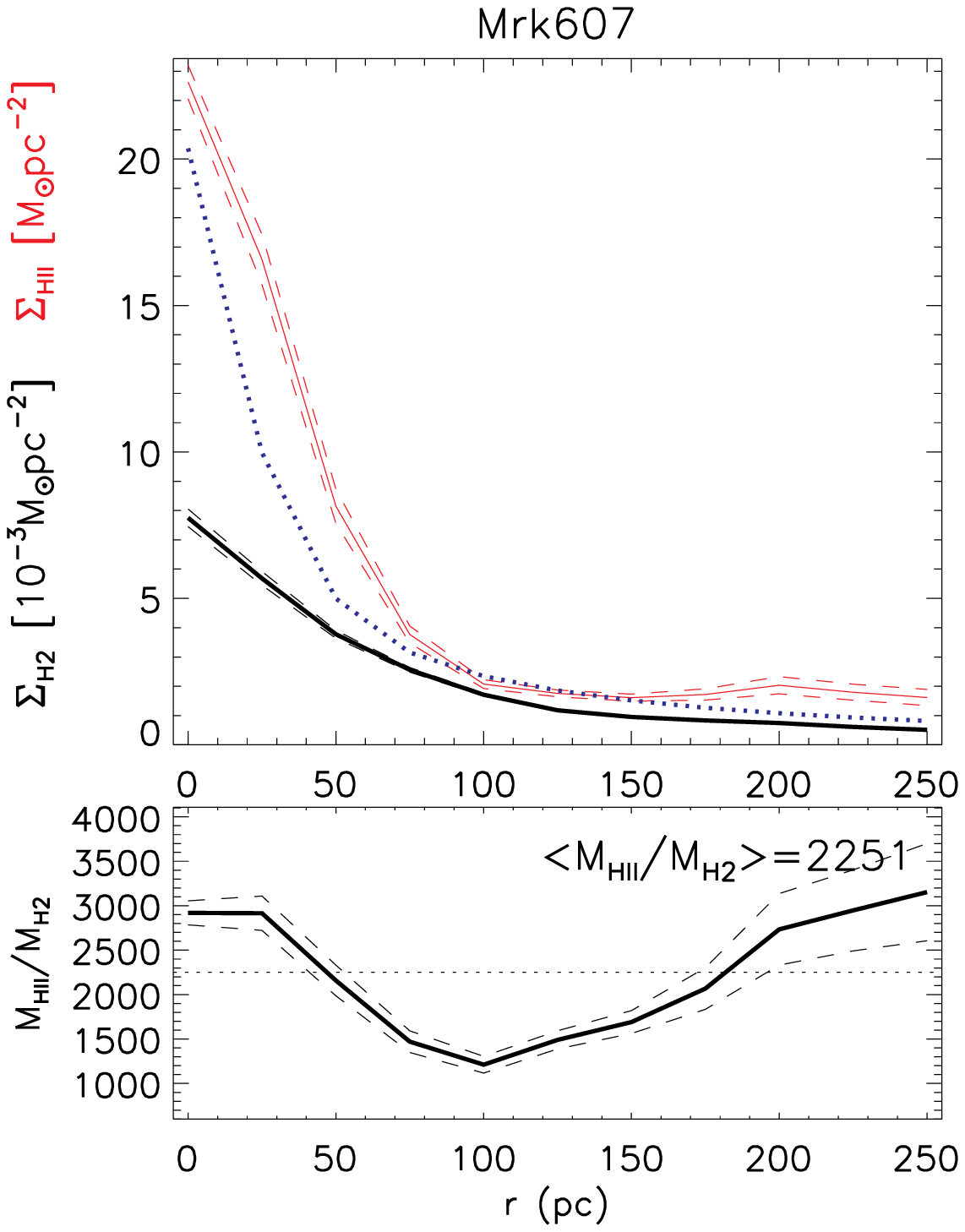}
\includegraphics[width=0.48\textwidth]{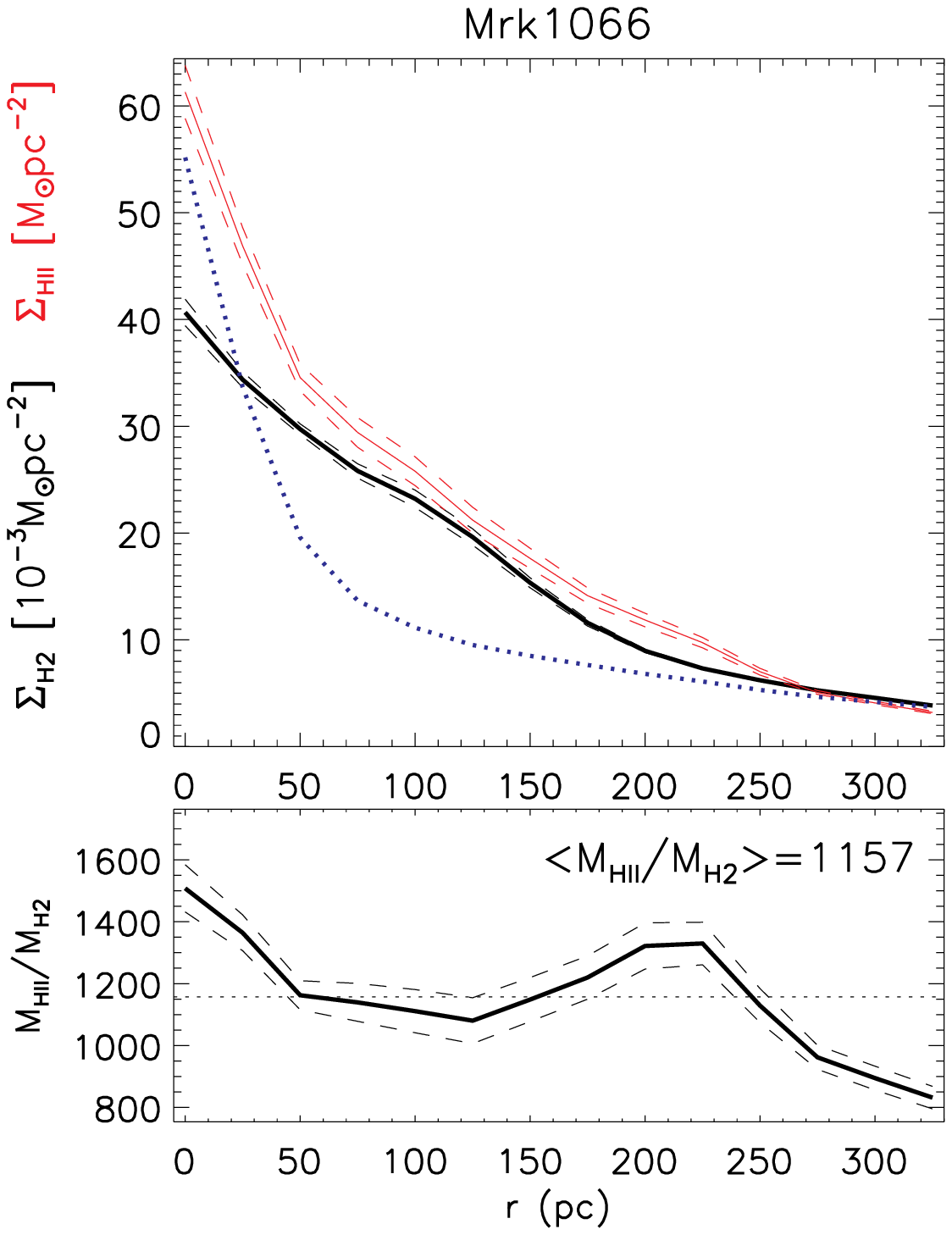}
\includegraphics[width=0.48\textwidth]{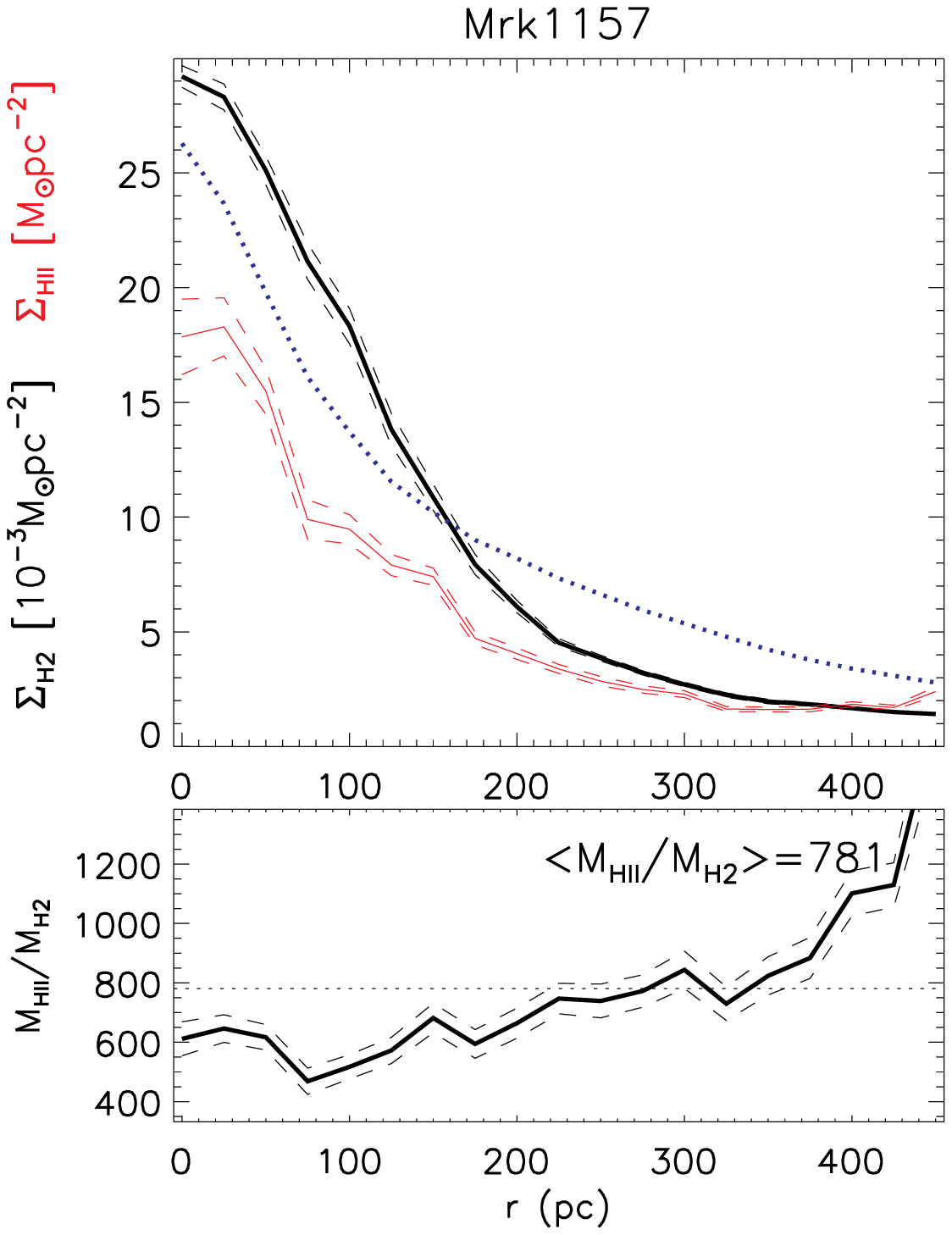}
\caption{Same as Fig.~\ref{sdenp} for Mrk\,607, Mrk\,1066 and Mrk\,1157.}
\label{sdenpd}
\end{figure*}

We used the equations \ref{mhii} and \ref{mh2} to calculate the molecular and ionized gas mass density spaxel-by-spaxel by  defining the gas surface mass densities of the molecular and ionized gas as  $\Sigma_{\rm H2}=\frac{M_{H_2}}{A_s}$ and $\Sigma_{\rm HII}=\frac{M_{\rm H\,II}}{A_s}$, respectively, where $A_s$ is the area of each spaxel. Using the calculated values of $\Sigma_{\rm H2}$ and $\Sigma_{\rm HII}$ we constructed the surface mass density profiles shown in Figures~\ref{sdenp}--\ref{sdenpd}.  Following \citet{barbosa06}, we calculated the position ($r$) of each spaxel in the plane of the disk as 
$r=\alpha R,$
where
\[ R=\sqrt{(x-x_0)^2+(y-y_0)^2}
\]
is the position projected in the plane of the sky (observed position)  and  
\[ \alpha=\sqrt{\rm cos^2(\Psi-\Psi_0)+sin^2(\Psi-\Psi_0)/cos^2(i)},
\]
where $\Psi_0$ is the orientation of the line of nodes, $i$ is the disk inclination and $\Psi={\rm tan^{-1}}\left(\frac{y-y_0}{x-x_0}\right)$ with ($x, y$) being the spaxel coordinates and ($x_0, y_0$) the location of the kinematical center.  Then, the surface mass density profiles were constructed by averaging the surface mass densities within concentric rings in the galaxy plane with width of $dr=$25 pc. For all galaxies we fixed the ($x_0, y_0$) as the position of the continuum peak and included only spaxels with flux measurements for the corresponding emission lines. For most galaxies, the H$_2\lambda2.12\,\mu$m and Br$\gamma$ flux maps have already been published by our group in the references listed in the last column of Table~\ref{sampleobs}. Although the \br\ line is weaker than \pb, its use is justified due to the fact that using \br\ and H$_2\,\lambda2.12$ lines, both ionized and molecular masses are derived from the same spectral band and thus the ratio between them is less sensitive to uncertainties in the flux calibrations and extinction, as both lines are are close in wavelength.   For two galaxies (NGC\,1052 and NGC\,5548), the \br\ line was not detected in our spectra and thus we used the \pb\ emission line to estimate $M_{\rm H\,II}$ by assuming the theoretical ratio between the fluxes of \pb\ and \br\ of 5.85 for the Case B  recombination \citep{osterbrock06}.   The references for the corresponding measurements as well as the discussion about the fitting procedures are listed in the last column of Table~\ref{gasdens}.  
This table presents also the adopted $\Phi_0$ and $i$ values, most of them from \citet{stel_llp}, who obtained these values by fitting the observed stellar velocity fields by rotation disk models and from the application of the technique of kinemetry to the measured kinematics. For Mrk\,3 and Mrk\,79 we used the disk geometric parameters from the Hyperleda database \citep{makarov14}, for NGC\,1068 from \citet{davies07} and for NGC\,4151 those presented in \citet{onken14}.

The top panels of figures~\ref{sdenp}--\ref{sdenpd} present for each galaxy the profiles for $\Sigma_{\rm H2}$ in black, in units of  $10^{-3}$\,M$_\odot$\,pc$^{-2}$, and $\Sigma_{\rm HII}$ in red in units of M$_\odot$\,pc$^{-2}$. 
The dotted blue line represents the K-band surface brightness  profile obtained from a continuum image derived by averaging the fluxes between 2.23 and 2.30\,$\mu$m. This profile is shown in units of $C\times$erg s$^{-1}$ cm$^{-2}$ \AA$^{-1}$ arcsec$^{-2}$ -- where $C$ is an arbitrary constant to put the profile in similar units to those of the mass density profiles -- to be used as a tracer of the stellar mass distribution.
The bottom panel shows the ratio between $\Sigma_{\rm HII}$ and $\Sigma_{\rm H2}$ or equivalently $\frac{M_{\rm H\,II}}{M_{H_2}}$, calculated considering only spaxels in which both \br\ and H$_2\lambda2.12$ flux measurements are available. The dotted horizontal line shows the mean value of $\frac{M_{\rm H\,II}}{M_{H_2}}$, indicated at the top-right corner of this panel and calculated from the $\Sigma_{\rm HII}/\Sigma_{\rm H2}$ profile. The dashed lines represent the standard error, calculated as the ratio between the standard deviation of the $\Sigma$ at each ring and the number of spaxels used to compute $\Sigma$. 

For all galaxies, the ionized and molecular gas mass density profiles decrease with the distance to the nucleus, with the ionized gas showing a steeper gradient for most galaxies. 
This behavior can be attributed to the different nature of the excitation mechanisms for the ionized and molecular gas: while the former is excited by the AGN radiation, the latter is dominated by thermal excitation through heating of the surrounding gas by X-rays emitted by the AGN \citep[e.g.][]{dors12,rogerio13,colina15}. As X-rays are less blocked by the surrounding gas, they penetrate in the disk more uniformly in all directions, so that the H$_2$ flux distributions are also more uniform than those of the ionized gas. The ionized gas usually shows more collimated flux distributions, as the AGN UV radiation is at least partially blocked by the dusty torus.
The only exception is NGC\,1068, that shows an increase in $\Sigma_{\rm H2}$ between 25 and 75~pc due to the presence of an expanding molecular gas ring \citep[e.g.][]{ms09,n1068-exc,barbosa14}. Both the ionized and molecular surface density profiles usually decrease more slowly with distance from the nucleus than the K-band brightness profile.  The fact that the gas mass density profiles are less steep than the stellar brightness profile is probably due to the fact that the gas has (more recently than the stars) settled in a disc, while the stellar density profile is dominated by stars from the galaxy bulge.  The bottom panels for each galaxy shows the radial profile for $\frac{M_{\rm H\,II}}{M_{H_2}}$, that confirm the trend that ionized gas shows an steeper decrease in surface mass density than the molecular gas, as the  $\frac{M_{\rm H\,II}}{M_{H_2}}$ for most galaxies have the highest values at the nucleus or at small distances from  it. The mean values of $<\frac{M_{\rm H\,II}}{M_{H_2}}>$, indicated at the top-left corner of each panel, range from $\sim$200 for Mrk\,607 to $\sim$8000 for NGC\,5506.

\begin{figure*}
\includegraphics[width=0.95\textwidth]{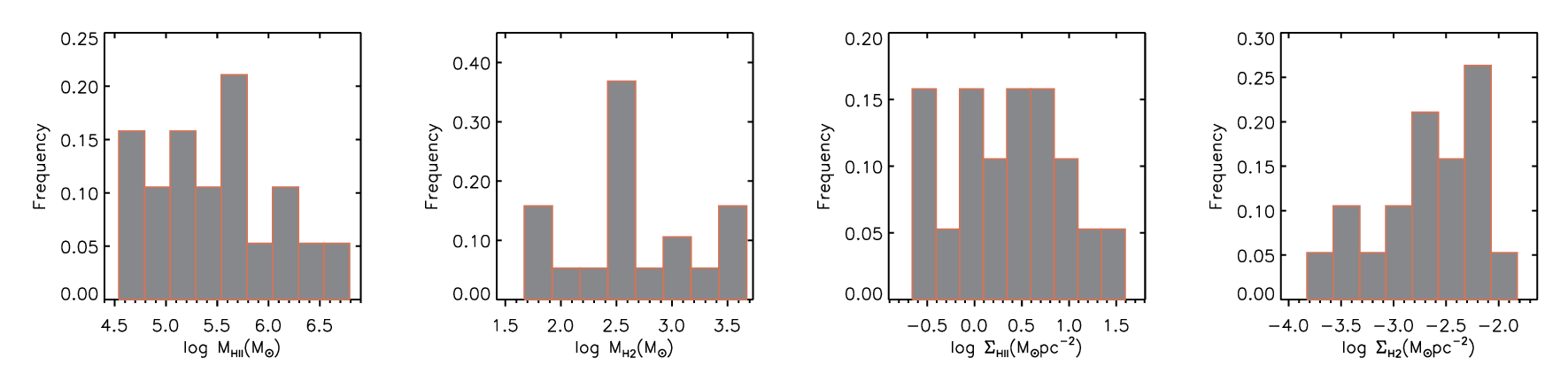}
\caption{Histograms for $M_{\rm HII}$, $\Sigma_{\rm HII}$, $M_{\rm H2}$ and $\Sigma_{\rm H2}$ for our sample, constructed using the values from Table~\ref{gasdens} using a bin of 0.25 dex.}
\label{histogram_prop}
\end{figure*}

Table~\ref{gasdens} shows the total mass of ionized and hot molecular gas for each galaxy by summing up the masses from all spaxels with detected \br\ and H$_2\lambda2.12\,\mu$m emission. The uncertainties in the masses are not included in this table, they are dominated by the uncertainty in flux calibration and can be up to 20\,\%. The mass of ionized gas is in the range $(3-440)\times10^4$\,M$_\odot$, while that for the hot molecular gas ranges from 50 to 3\,000 M$_\odot$. The mean surface mass density for the ionized and molecular gas, shown in Table~\ref{gasdens} are in the ranges (0.2--35.9)\,M$_\odot$\,pc$^{-2}$ and (0.2--13.9)$\times10^{-3}$\,M$_\odot$\,pc$^{-2}$. These values are in good agreement with those previously obtained, summarized by \citet{n5548} in their Table 1.  The distribution of ionized and molecular masses and surface mass densities for the galaxies of our sample are presented in Figure~\ref{histogram_prop}.

\begin{figure}
\includegraphics[width=0.45\textwidth]{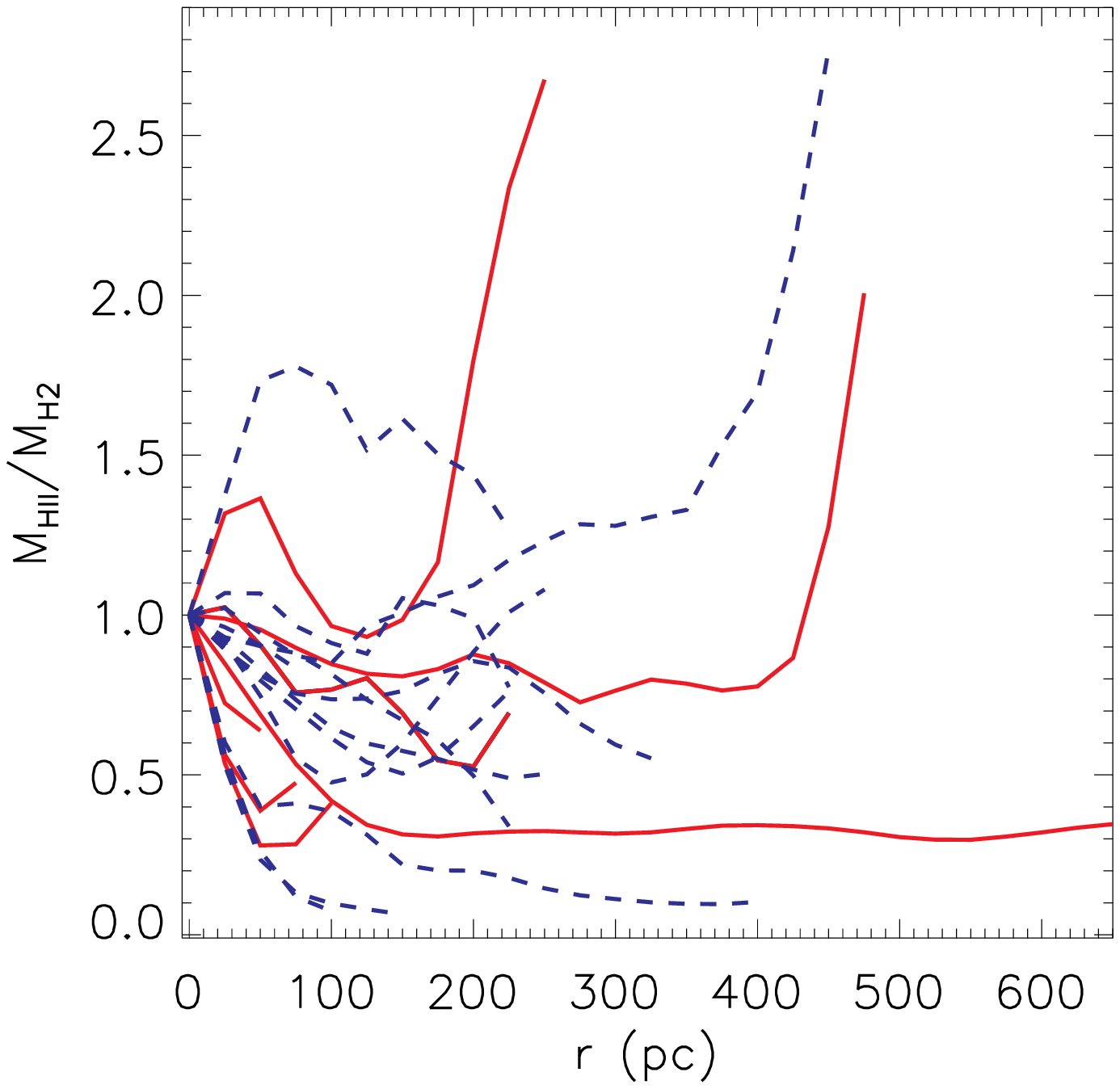}
\caption{Normalized  (by the nuclear value) radial profiles of $M_{\rm HII}/M_{\rm H2}$. Seyfert 1 galaxies are shown as continuous red lines and Seyfert 2 galaxies as dashed blue lines.}
\label{ratiodens}
\end{figure}

In order to further investigate the distribution of ionized and molecular gas in the inner few hundreds of parsercs of the galaxies of our sample, we constructed normalized radial profiles by dividing the $M_{\rm HII}/M_{\rm H2}$ value at each radial bin by the nuclear value ($r<25$\,pc). These profiles are shown in Figure~\ref{ratiodens}. Seyfert 1 galaxies (Mrk\,766, Mrk\,79, NGC\,3227, NGC\,3516, NGC\,4051, NGC\,4151 and NGC\,5548) are shown as red continuous lines and Seyfert 2 galaxies (Mrk\,1066, Mrk\,1157, Mrk\,3, Mrk\,607, NGC\,1052, NGC\,1068, NGC\,2110, NGC\,4388, NGC\,5506, NGC\,5899, NGC\,5929, NGC\,788) as blue dashed lines. These profiles confirm the result already mentioned above that the ionized gas has an steeper surface mass profile, as for most galaxies the  $M_{\rm HII}/M_{\rm H2}$ decreases with the distance to the nucleus. In addition, Figure~\ref{ratiodens} shows that there is no significant difference for the distribution of ionized and molecular gas for Seyfert 1 and Seyfert 2 nuclei.

\begin{table*}
\scriptsize 
\caption{Molecular and ionized gas masses and surface densities. (1) Name of the galaxy; (2) Total mass of ionized gas; (3) Area for the Br$\gamma$ emission; (4) Average surface mass density for the ionized gas; (5) Total mass of hot molecular gas; (6) Area for the H$_2\,\lambda 2.12$ emission; (7) Average surface mass density for the hot molecular gas; (8) Average star formation density; (9) total star formation rate; (10) and (11) orientation of the major axis and inclination of the disk, used in the deprojection from \citet{stel_llp}, except for Mrk\,3 and Mrk\,79 \citep[from Hyperleda database][]{makarov14}, NGC\,1068 \citep[from][]{davies07} and NGC\,4151 \citep[from][]{onken14}; (12) AGN bolometric luminosity estimated from the 14-195 keV luminosity; (13) mass accretion rate onto the SMBH; (14) Reference for the  H$_2\,\lambda 2.12$ and Br$\gamma$ flux maps.}
\vspace{0.3cm}
\begin{tabular}{l c c c c c c c c c c c c c }
\hline
       (1)         &   (2)     &    (3)            &  (4)   &    (5)    & (6)	     & (7) & (8) & (9) &(10) &  (11) & (12) & (13) & (14) \\
    Galaxy  & $M_{\rm HII}$ & $A_{\rm HII}$ & $<\Sigma_{\rm HII}>$ & $M_{\rm H2}$ & $A_{\rm H2}$ & $<\Sigma_{\rm H2}>$  & $<\Sigma_{\rm SFR}>$  &  $SFR$ & $\Psi_0$  & $i$ & log\,$L_{\rm bol}$ & $\dot{m}$ & Ref.\\
        &$10^4 {\rm M_\odot}$ &  $10^4 {\rm pc^2}$ & ${\rm M_\odot/pc^2}$ &$10^2{\rm M_\odot}$ &$10^4 {\rm pc^2}$ & ${\rm 10^{-3} M_\odot/pc^2}$ &${\rm 10^{-3} M_\odot/yr\,kpc^2}$ & ${\rm 10^{-4} M_\odot/yr}$ & deg & deg &  erg/s & $10^{-3}$\,M$_\odot$/yr & \\
\hline
\multicolumn{14}{c}{Main Sample} \\
\hline
   NGC788   &  36.11  &   8.30  &   4.35  &  10.90  &  70.25  &   1.55  &   1.96  & 1.63 &120  &   20.8 & 44.4 & 49.2 & a\\
   NGC1068  &   3.47  &  15.59  &   0.22  &   0.54  &  15.86  &   0.34  &   0.03  & 0.05 & 145 &   40.0 & 42.8 &  1.1 & b \\
   NGC2110  &  28.17  &  11.00  &   2.56  &  15.66  &  23.79  &   6.59  &   0.93  & 1.03 & 156 &   42.5 & 44.6 & 65.4 & c \\
      Mrk3  &  22.39  &  59.90  &   0.37  &   0.47  &  31.03  &   0.15  &   0.06  & 0.38 & 15  &   31.7 & 44.7 & 87.2 & d \\
   NGC3227  &  17.87  &   2.69  &   6.64  &   7.81  &   5.95  &  13.12  &   3.54  & 0.95 & 156 &   45.4 & 43.4 &  4.1 & a \\
   NGC3516  &  11.62  &  15.94  &   0.73  &   3.87  &  30.63  &   1.26  &   0.16  & 0.26 &  54 &   12.8 & 44.2 & 27.9 & a \\
   NGC4151  &  59.22  &   5.93  &   9.99  &   2.87  &   5.72  &   5.02  &   6.27  & 3.72 & 85  &   23.0 & 44.0 & 16.0 & e \\
    Mrk766  &  51.60  &   2.93  &  17.60  &   3.75  &   6.53  &   5.74  &  13.85  & 4.06 & 66  &   18.2 & 44.0 & 16.0 & f \\
   NGC4388  &   4.36  &  18.37  &   0.24  &   0.67  &  22.31  &   0.30  &   0.03  & 0.06 & 96  &   27.7 & 44.6 & 65.4 & a \\
   NGC5506  & 439.64  &  12.26  &  35.86  &   8.89  &  12.00  &   7.41  &  37.53  & 46.01&  96 &   58.7 & 44.3 & 37.0 & a \\
\hline
\multicolumn{14}{c}{Complementary Sample} \\
\hline
   NGC1052  &   6.18  &   0.92  &   6.70  &   3.00  &   7.56  &   3.96  &   3.58  & 0.33 & 114 &   47.5 & 42.9  & 1.4 & g \\
   NGC4051  &   3.64  &   2.13  &   1.71  &   0.86  &   2.91  &   2.94  &   0.53  & 0.11 & 24  &   37.3 & 42.5  & 0.5 & h \\
   NGC5548  &  74.49  &  85.30  &   0.87  &   3.80  &  15.70  &   2.42  &   0.21  & 1.76 & 108 &   60.9 & 44.7  & 87.2 & i \\
   NGC5899  &   8.10  &   3.81  &   2.12  &   3.60  &  23.45  &   1.53  &   0.72  & 0.27 & 24  &   62.7 & 43.1  & 2.4  & a \\
   NGC5929  &  14.70  &  23.77  &   0.62  &   3.94  &  28.86  &   1.37  &   0.13  & 0.30 & 30  &   60.7 & --  & --  & j \\
     Mrk79  & 169.24  & 163.24  &   1.04  &  26.90  & 179.32  &   1.50  &   0.26  & 4.29 & 73  &   35.6 & 44.8  &116.4 & k \\
    Mrk607  &  51.85  &  20.30  &   2.55  &   2.06  &  28.49  &   0.72  &   0.93  & 1.89 & 138 &   58.2 & --  &  --  & a \\
   Mrk1066  & 305.89  &  31.45  &   9.73  &  30.11  &  45.02  &   6.69  &   6.04  & 19.0 & 120 &   50.2 & --  &  -- & l \\
   Mrk1157  & 188.70  &  65.13  &   2.90  &  28.24  &  89.60  &   3.15  &   1.11  & 7.22 & 114 &   45.1 & --  & --  & m \\
\hline
\multicolumn{14}{l}{a: Schonell et al., in prep.; b: \citet{n1068-exc}; c: \citet{diniz15} ; d: Fischer et al., in prep.; e: \citet{sb09} }\\
\multicolumn{14}{l}{f: \citet{mrk766}; g: Dahmer-Hahn et al, in prep.;  h: \citet{n4051}; i: \citet{n5548}; j: \citet{n5929};  }\\
\multicolumn{14}{l}{ k: \citet{mrk79}; l: \citet{mrk1066-exc}; m: \citet{mrk1157}}\\

\hline
\end{tabular}
  \label{gasdens}
\end{table*}


\section{Feeding the AGN and Star Formation} \label{feeding}

We can estimate the accretion rate ($\dot{m}$) to the AGN in each galaxy by

\begin{equation}
 \dot{m}=\frac{L_{\rm bol}}{c^2\eta},
\end{equation}
where $L_{\rm bol}$ is the AGN bolometric luminosity, $c$ 
is the light speed  and $\eta$ is the efficiency 
of conversion of the rest mass energy of the accreted material into radiation.  The AGN bolometric luminosity can be estimated from the hard X-ray luminosity by \citep{ichikawa17} 
\begin{equation}
{\rm log} L_{\rm bol} = 0.0378({\rm log} L_{\rm X})^2 - 2.03{\rm log} L_{\rm X} +61.6,
\end{equation}
where $L_X$ is the hard X-ray (14-195 keV) luminosity. The resulting $\dot{m}$ values using the X-ray luminosities from  Table~\ref{sampletab} and assuming  $\eta\approx0.1$ \citep[e.g.][]{frank02}, are shown in Table~\ref{gasdens}. The $\dot{m}$ for our sample ranges from 10$^{-4}$ (for NGC\,4051) to 10$^{-1}$ M$_\odot$ yr$^{-1}$ (for Mrk~79), with a mean value of $<\dot{m}>\sim0.03$~M$_\odot$ yr$^{-1}$. 

The surface mass density profiles of Figs.~\ref{sdenp}--\ref{sdenpd} show that most of the ionized and molecular gas masses listed in Table~\ref{gasdens} are concentrated within the inner $\sim$300 pc of the galaxies. The ionized gas mass alone would be enough to feed the central AGN for an activity cycle of $10^7-10^8$ yr. The hot molecular gas mass is typically 3 orders of magnitudes lower than that of the ionized gas, but this gas is just the heated surface of a probably much larger molecular gas reservoir of colder molecular gas, that may be 10$^5 - 10^7$ times more massive \citep{dale05,ms06,mazzalay13}, implying that the masses of the cold molecular gas probably range from 10$^7 - 10^9$\,M$_\odot$.


We conclude that, within the inner 300 pc of our sample, there is at least  $\sim10^2$ times more gaseous mass than the necessary to feed the AGN. Most of this mass will not feed the AGN and might be consumed by star formation. 
The pioneer work by \citet{schmidt59} showed that the star formation rate ($SFR$) is directly related to the gas density,  while \citet{kennicutt98} derived a relation between the SFR surface density ($\Sigma_{SFR}$) and the ionized gas mass surface density ($\Sigma_{\rm HII}$) so that the former can be obtained from the latter as

\begin{equation}
\frac{\Sigma_{SFR}}{M_\odot yr^{-1} kpc^{-2}} = (2.5\pm0.7)\times 10^{-4} \left(\frac{\Sigma_{\rm HII}}{M_\odot pc^{-2}} \right)^{1.4},
\label{eqsfr}
\end{equation}
where $\Sigma_{\rm HII}$ is the surface mass density.

Using the equation above, we obtained the mean values of the star formation density <$\Sigma_{SFR}$>  for each galaxy, shown in Table~\ref{gasdens}, which varies from $3\times10^{-5}$ to $3.8\times10^{-2}$  $M_\odot yr^{-1} kpc^{-2}$. We point out that these should be minimum values, as we are considering only the ionized gas, and there should much more molecular gas than traced by the hot molecular gas phase that we have observed. Considering the area of the \br\ emission quoted in Table~\ref{gasdens}, we obtain a wide range of minimum total star formation rate of 10$^{-6}$--10$^{-3}$ M$_\odot$\,yr$^{-1}$ (shown in Table~\ref{gasdens}).   These values of $SFR$ are smaller 
smaller than those usually obtained for the nucleus of star-forming galaxies and circumnuclear rings of star formation ($SFR\sim10^{-3}$\,M$_\odot$\,yr$^{-1}$) \citep[e.g.][]{wold06,shi06,dors08,galliano08,falcon-barroso14,n4303}. Considering a scenario in which the total mass would be used to form stars, the estimated masses for our sample would allow the star formation for about 10$^9$ yr at the current star formation rate.

Thus, considering the derived mass accretion rate, the star formation rate and the mass of molecular and ionized gas, we conclude that the mass reservoirs of the galaxies of our sample are much larger than that needed to power the central AGN and star formation, thus allowing the co-existence of recent star formation \citep[as evidenced by low-stellar velocity dispersion structures seen in some galaxies,][]{stel_llp} and the nuclear activity.

\section{Conclusions}\label{conclusions}

We characterized a sample of 20  nearby X-ray selected Seyfert galaxies being observed with the NIFS instrument of the Gemini North Telescope plus a complementary sample of 9 additional galaxies already observed with NIFS. We also present and discuss mean radial profiles within the inner kiloparsec for the ionized and molecular gas surface mass densities for the galaxies already observed: 11 from the main X-ray sample and 9 galaxies from the  complementary sample. Our main conclusions are:

\begin{itemize}

\item The average values of X-ray luminosities are $<{\rm log} L_{\rm X}> =42.6\,\pm\,0.1$ erg\,s$^{-1}$ for the main sample and $<{\rm log} L_{\rm X}> =42.4\,\pm\,0.1$ erg\,s$^{-1}$ for the main plus complementary sample. The \oiiil\ luminosities are in the range  $L_{[OIII]} = (0.2-155)\times10^{40}$\,erg\,s$^{-1}$, with a mean value of $<{\rm log} L_{\rm [OIII]}> =41.0\,\pm\,0.2$ erg\,s$^{-1}$.

\item The $M_B$ and $M_H$ distributions for the restricted BAT sample (all galaxies with  $L_X \ge 10^{41.5}$ ergs\,s$^{-1}$ and $z\le$0.015 from the 60 month BAT catalogue)  and our sample are very similar, indicating that the additional criteria used in the definition of our sample does not include any bias in terms of these properties. The mean values for our sample are $<M_B>=-20.75\pm0.16$ and $<M_H>=-23.83\pm0.13$.

\item The mean value of the central stellar velocity dispersion of the total sample is 154$\pm$11\,\kms, being essentially the same as that of the X-ray sample only.

\item The axial ratio $b/a$ of the total sample 
ranges from 0.2 (corresponding to a disk inclination of $i\sim 80^\circ$, almost edge-on) to 0.9 ($i\sim25^\circ$, almost face-on).

\item We constructed mean radial profiles for the surface mass density of the ionized ($\Sigma_{\rm HII}$) and hot molecular ($\Sigma_{\rm H2}$) gas for the 20 galaxies already observed, derived from the \br\ and \hml\ fluxes.  Both profiles decrease with the distance from the nucleus for most galaxies, with the ionized gas showing a steeper gradient. The only exception is NGC\,1068, which shows an increase in  $\Sigma_{\rm H2}$ at 25--75~pc from the nucleus due to the presence of a molecular gas ring. We attribute this difference in behavior to the distinct origin of the gas emission: while for the H$^+$ the emission is due to recombination of ionized gas by the AGN, for the H$_2$ the excitation is mostly thermal due to the heating of the gas by X-rays that penetrate deeper into the surrounding gas in the galaxy plane.

\item The mean surface mass density for the ionized and molecular gas are in the ranges (0.2--35.9)\,M$_\odot$\,pc$^{-2}$ and (0.2--13.9)$\times10^{-3}$\,M$_\odot$\,pc$^{-2}$, respectively, while the ratio between them ranges from $\sim$200 for Mrk\,607 to $\sim$8000 for NGC\,5506. The mean star formation surface density is $<\Sigma_{SFR}>=(4.09\pm0.44)\times{\rm 10^{-3} M_\odot\,yr^{-1}\,kpc^{-2}}$, while the star formation rates range from  10$^{-6}$ to 10$^{-3}$ M$_\odot$\,yr$^{-1}$, with the accretion rate onto the SMBH ranging from 10$^{-4}$ to 10$^{-1}$ M$_\odot$ yr$^{-1}$. 

\item The total mass of ionized gas within the inner $\sim100-500$\,pc is in the range $(3-440)\times10^4\,{\rm M_\odot}$, while that of hot molecular gas ranges between 50 and 3000 ${\rm M_\odot}$. Considering also that the mass of cold molecular gas is usually $\sim10^5$ times larger than that of hot molecular gas for AGN in general, we estimate a total mass of gas in the region ranging from 10$^6$ to 10$^8~ {\rm M_\odot}$. Comparing these masses with the typical accretion rates above, it can be concluded that they are much larger than that necessary to feed a typical AGN cycle of of $10^7-10^8$ yr. The fate of this gas is probably the formation of new stars in the region (the AGN-Starburst connection).


\end{itemize}

\section*{Acknowledgments}
We thank an anonymous referee for useful suggestions which helped to improve the paper. 
Based on observations obtained at the Gemini Observatory, 
which is operated by the Association of Universities for Research in Astronomy, Inc., under a cooperative agreement with the 
NSF on behalf of the Gemini partnership: the National Science Foundation (United States), the Science and Technology 
Facilities Council (United Kingdom), the National Research Council (Canada), CONICYT (Chile), the Australian Research 
Council (Australia), Minist\'erio da Ci\^encia e Tecnologia (Brazil) and south-eastCYT (Argentina).  
This research has made use of the NASA/IPAC Extragalactic Database (NED) which is operated by the Jet
 Propulsion Laboratory, California Institute of  Technology, under contract with the National Aeronautics and Space Administration.
We acknowledge the usage of the HyperLeda database (http://leda.univ-lyon1.fr).
This publication makes use of data products from the Two Micron All Sky Survey, which is a joint project of the University of Massachusetts and the Infrared Processing and Analysis Center/California Institute of Technology, funded by the National Aeronautics and Space Administration and the National Science Foundation.
The Brazilian authors acknowledge support from FAPERGS,  CNPq and CAPES.
L.B. was partly supported by a DFG grant within the SPP 1573 ``Physics of the interstellar medium''.










\bsp	
\label{lastpage}
\end{document}